\documentclass[%
reprint,
 amsmath,amssymb,
 aps,
	]{revtex4-2}

\usepackage{graphicx}
\usepackage{dcolumn}
\usepackage{bm}
   \usepackage{xcolor}
   \definecolor{best_acc}{rgb}{0,0.5,0} 
   \definecolor{refcol}{rgb}{0,0,1} 
   \definecolor{mygray}{gray}{0.85} 
   \definecolor{mycite}{rgb}{0,0,1} 
   \definecolor{edit1}{rgb}{1,0,0}
   \definecolor{edit2}{rgb}{0.62,.12,.941}
   \definecolor{edit1}{rgb}{0,0,0}
   \definecolor{edit2}{rgb}{0,0,0}
   \definecolor{myurl}{rgb}{0,.45,0.85} 
   \usepackage[colorlinks=blue]{hyperref}
   \hypersetup{
   	colorlinks=false,  
   	pdfborder={0 0 0}, 
   }
   \hypersetup{
   	colorlinks=true,
   	linkcolor=refcol,
   	citecolor= mycite,
   	filecolor=magenta,      
   	urlcolor=myurl,
   }
\graphicspath{{./figures/}{../figures/}}

\usepackage{lineno}

\begin{document}
	\title{Topological Protection in a Strongly Nonlinear Interface Lattice}
	\author{Joshua R.~Tempelman}
	\author{Kathryn H.~Matlack}
	\author{Alexander F.~Vakakis}
	\affiliation{%
		Department of Mechanical Science and Engineering, University of Illinois at Urbana-Champaign, 1206 W Green St, Urbana, IL 61801
	}%

\begin{abstract}
		Mechanical topological insulators are well understood for linear and weakly nonlinear systems, however traditional analysis methods break down for  strongly nonlinear systems since linear methods can not be applied in that case. We study one such system in the form of a one-dimensional mechanical analog of the Su-Schrieffer-Heeger interface model with strong nonlinearity of the cubic form. 
		The frequency-energy dependence of the nonlinear bulk modes and topologically insulated mode is explored using Numerical continuation of the system's nonlinear normal modes (NNMs), and the linear stability of the NNMs are investigated using Floquet Multipliers (FMs) and Krein signature analysis. 
		We find that the nonlinear topological lattice supports a family of topologically insulated NNMs that are parameterized by the total energy of the system and are stable within a range of frequencies. 
		\textcolor{edit2}{
		Next, it is shown that empirical calculations of the geometric Zak Phase can define an energy threshold to predict the excitability of the nonlinear topological mode, and that this threshold coincides with the energy that the topological NNM intersects the linear bulk-spectrum. These predictions are validated with numerical simulations of the nonlinear topological system.}
		These results are also tested for parametric perturbations which preserve and break chirality in the system.
		\textcolor{edit2}{
		Thus, we provide a new method for analyzing and predicting the existence of topologically insulated modes in a strongly nonlinear lattice based on the physical observable of band topology.
		}
\end{abstract}

	\maketitle
	
\section{Introduction}
Recent advances in the field of condensed matter physics have unlocked new tools for engineering materials with exotic band properties. 
Namely, the nontrivial band topology existent in certain quantum-mechanical systems has enabled the conception of topological insulators~\cite{Hasan2010}.   
In the framework of mechanical and acoustic wave control, topological systems offer unique and unprecedented wave propagation characteristics such as a robustness to perturbations and immunity to back scattering. 
Topological insulators allow for a wave which is isolated (and ``protected'' against perturbations in the ``bulk'' of a material) at an edge or interface; it's existence is established by the topological properties of the bulk-bands thus making it a robust mechanism for isolating and directing energy~\cite{Chaunsali2017,Chaunsali2021}.
This is the working principle in bulk-boundary correspondence, that is, a phenomenon that allows for predication of waves at band edges based on an invariant of the bulk~\cite{Essin2011,Chaunsali2019}. 
In one-dimensional (1D) lattices, the insulated wave manifests at an edge or interface. 
For two and three-dimensional insulators (2D and 3D), the topological wave can be captured on edges, corners, or surfaces, depending on the nature of the insulator~\cite{Smirnova2020}. 

The theories supporting topological insulation were originally envisioned for the quantum domain, however topological insulation has been shown to have macroscopic analogs in systems based in classical physics as well~\cite{Suesstrunk2016}. This has enabled researchers to extrapolate this exotic behavior to many forms of waves.
For example, 1D topological insulators have already displayed efficacy in acoustic and elastic phononics and metamaterials~\cite{Xiao2015,Ma2019,Zhao2018, 
	Pal2019, 
	Yin2018,Arretche2020, Chaunsali2019,Pal2018,Liu2019}. 
Beyond being a fascinating topic of study in its own right, the robust wave control properties that topological insulators inflict in these physical domains make them attractive candidates for applications such as energy harvesting, vibration control, and wave-based information transfer~\cite{Darabi2020}. 																			

The geometric phase conditions that a topological insulator must satisfy are constructs of system's linear dynamics. However,		 nonlinearity can often manifest in mechanical systems and consequently render the linear analysis of little use; the proper treatment of topological insulators with strongly nonlinear elements remains an open question.													
The study of nonlinear topological insulators has therefore gained traction across both the physics and mechanics					 communities~\cite{Smirnova2020}. 																									
Recent work on 1D and 2D topological lattices with weak nonlinearity employed asymptotic techniques to show analytically that a nonlinearity of the (hardening or softening) cubic form can shift the topological mode into the bulk-bands~\cite{Pal2018}. 	
\textcolor{edit2}{	
Nonlinearity in a topological lattice has also been explored using an effective stiffness approach to show that contact	nonlinearity 
can induce topological transitions in both analytical and experiential frameworks~\cite{Chaunsali2017,Chaunsali2019}; this is based on the fact that precompressed granular systems can demonstrate effective linear properties~\cite{Starosvetsky2017}. 	}			
%
%
%
Furthermore, the existence of nonlinearities in topological insulators has been used to construct protected solitons in lattices~\cite{Leykam2016,Snee2019,Lumer2013} and study the stability of nonlinear quantum walks~\cite{Gerasimenko2016,Mochizuki2020}. 	

The recent thrusts in nonlinear topological insulators largely ignore the strongly nonlinear regime. 
When strong nonlinearity is assumed, asymptotic methods based, e.g.,~on linearized generating solution, are inapplicable and new ways of studying the nonlinear dynamics must be explored. 
Thus far, the majority of work on strongly nonlinear insulators does so in the quantum framework whereby first order equations dictate the system dynamics~\cite{Chaunsali2021,Zhou2020}. 
In phononics and elastic media, second order differential equations govern the system dynamics. 
\textcolor{edit1}{
Exploratory work to study the strongly nonlinear topological lattice with second order dynamics was explored first by Vila \textit{et.~al.}~\cite{Vila2019} and more recently by Chaunsali \textit{et.~al.}~\cite{Chaunsali2021} whereby both studies use numerical continuation of the nonlinear normal modes (NNMs) for a mechanical analog of the Su-Sheiffer-Heeger (SSH) model with grounding stiffness. 
}
\textcolor{edit2}{
Both~\cite{Chaunsali2021,Vila2019} relate either experimental or numerical findings to the frequency-energy profiles uncovered in continuation analysis, and~\cite{Chaunsali2021} employed Floquet stability theory to predict the linear stability of nonlinear topological edge modes.
} 
\textcolor{edit1}{
}

\textcolor{edit2}{
While NNMs are an effective tool for recovering modes of periodic oscillation at high energies and developing the frequency-energy evolution of the system, they are only valid for oscillations which satisfy the NNM profile perfectly. In practice, the topological modes must be excited from external forces in a similar fashion to the experiments described in~\cite{Vila2019,Chaunsali2017};
the criteria for predicting at which energies the mode is excitable is not well understood.
%
Additionally, previous connections between experimental findings and the system's NNMs have in-large refrained from relating the observations to topological band theory.}
%
\textcolor{edit2}{
In addition, prior work has observed that a \textit{merging} of the nonlinear topological NNM with linear bulk modes leads to the disincorporation of the topological mode shape~\cite{Chaunsali2021,Vila2019}.
Naturally, this leads to the question of whether empirical calculations of the Zak phase is linked to the continuation of the topological mode and bulk-bands. Moreover, it is known that the bulk bands of nonlinear lattices evolve with energy ~\cite{Jayaprakash2010,Mojahed2019,Mojahed2019a}, and the relation of nonlinear bulk-band evolution with evolution of the topological mode has yet to be investigated. We will explore this issue among others in this work.
}

\textcolor{edit1}{
In this work, we explore the influence of strong nonlinearity in a mechanical lattice analogous to the SSH model. 
A strong nonlinearity is introduced in the stiff inter-cell coupling of a diatomic chain, and a grounding term is added for each oscillator. 
\textcolor{edit2}{
The system is first studied by performing numerical continuations and employing Floquet theory to predict the linear stability of the corresponding nonlinear standing wave solutions during the frequency-energy evolution of the NNMs. 
This methodology is used to reveal the evolution and stability of the nonlinear system's topological mode and bulk modes, and this
 is graphically demonstrated through the system's frequency-energy plot (FEP).
}
Additionally, we use the NNMs to compare the treatment of strongly nonlinear topological systems in previous studies to methods developed in our own work.
The geometric phases of the lattices are studied in numerical experiments to define a critical energy threshold below which the required phase condition for topological insulation exists in the optical band of the nontrivial lattice.
 \textcolor{edit2}{
We find that this energy threshold is located approximately at the same energy level that the NNM of the topological mode penetrates the linear optical band.
Thus, our approach can be seen as an alternative method which is based on the observable characteristics of band-topology present in the transient dynamics of the bulk modes.
}
  Lastly, full-system simulations are studied by exciting the interface mode with Gaussian tone-bursts, and it is shown that the critical energy threshold defined in the empirical phase experiments are successful in predicting the energy levels at which the topological mode can be excited. 
}

The outline of this paper is as follows. Section~\ref{sec:NumericalContinuation and Nonlianer Pass Bands} describes the linear properties of the topological interface lattice that is studied in this work, as well as the numerical continuation of the nonlinear system to study the nonlinear topologically insulated mode and the nonlinear bulk-bands. Section~\ref{sec: Stability Analysis} studies the linear stability of NNMs of the nonlinear topological lattice by use of Floquet Multipliers and Krein signatures. Section~\ref{sec:Geomteric Phase} discusses the extension of the system's linear topological invariant in the strongly nonlinear regime by use of numerical experiments. The usefulness of this phase analysis in predicting the existence of the nonlinear topological mode at moderate-to-high energy levels, the effect of forcing profiles on the system dynamics, and the effect of perturbations on the topological mode are discussed in section~\ref{sec:Full System Analysis}, whereas section~\ref{sec: conclusion} offers some concluding remarks.

\section{Linear Analysis and Nonlinear Continuation}
\label{sec:NumericalContinuation and Nonlianer Pass Bands}
We start by studying the linear lattice to confirm that the analytical requirements for a topological insulator are satisfied in the low-energy limit.
Next, the goal is to study the behavior of the insulated mode and bulk-modes as the nonlinearity becomes increasingly prevalent (i.e.,~as the energy of the system is increased).
In strongly nonlinear systems, it is seldom possible to produce analytical expressions for the dynamics at high energy states. 
Therefore, we use a \textit{shooting method}, to numerically resolve time-periodic solutions at various frequencies and energies. These produce the nonlinear normal modes (NNMs) of the topological mode and the bulk-modes, and these will be the basis for the frequency energy plot (FEP) which will be constructed to describe the evolution of the NNMs (standing wave solutions).

\begin{figure*}[t!]
	\centering
	\includegraphics[width=\linewidth]{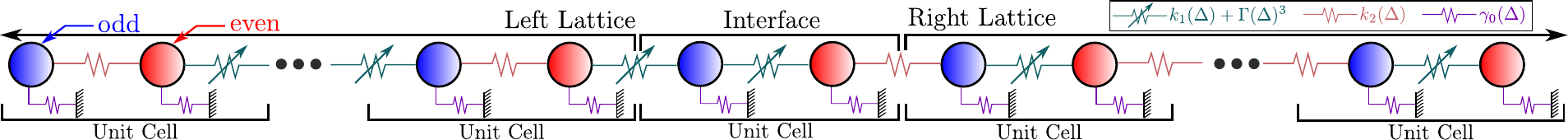}
	\caption{Schematic of the topological interface lattice.}
	\label{Fig:Schematic}
\end{figure*}

\subsection{The Linear System and Topological Invariant}
\label{sec: Linaer System}
Our system draws inspiration from the SSH model with onsite potential in the form of a 1D insulator with a well-studied classical analog~\cite{Hasan2010,Chaunsali2021,Pal2018,Asboth2016}. 
A diatomic chain is constructed of oscillators connected by coupling springs (Figure~\ref{Fig:Schematic}), as well as grounding springs. The intra-coupling springs take one of two values, $k_1$ or $k_2$, which are parameterized by a scalar $\gamma$; namely $k_1 = k(1+\gamma)$ and $k_2 = k(1-\gamma)$, with $k=1$ and $\gamma = 0.4$. The linear grounding stiffness is uniform with characteristic stiffness of $\gamma_0 = 1$. 
A nonlinear coupling term is parameterized by $\Gamma$ with nominal value $\Gamma = 0.5$.

The governing equations of the lattice are separated into three regimes, that is, the the left of the interface, to the right of the interface, and at the interface, respectively. 
%
%
%
%
%
%
\textcolor{edit2}{
The equations of motion can be written in the compact form
\begin{equation}
	\ddot{\bm{u}}+\textbf{K}\bm{u} + \bm{f}_{\rm NL} = \textbf{0}
\end{equation}
where $\bm{u}$ is the displacement vector,
$\bm{f}_{{\rm NL},i}$ influences inter-cell couplings for cells to the left of the interface, e.g.~$\bm{f}_{{\rm NL},i} = \Gamma(u_{e,i} -u_{o,i+1})^3$, whereas  $\bm{f}_{{\rm NL},i}$ influences intra-cell couplings for cells to the right of the interface,  $\bm{f}_{{\rm NL},i} = \Gamma(u_{e,i} - u_{o,i})^3$ where subscripts $(\cdot)_{o}$ denote  odd lattice sites  and subscripts $(\cdot)_{e}$ denote even lattice sites. The matrix $\textbf{K}$,
 can be split into the left, interface, and right components ($\textbf{K}_L$, $\textbf{K}_I$,
  $\textbf{K}_R$),
 \[\textbf{K} = \begin{bmatrix}
 	\textbf{K}_L &\textbf{K}^*_L & \textbf{0} & \cdots& \textbf{0}\\
 	\textbf{0} &\cdots & \textbf{K}_I & \cdots& \textbf{0}\\
 	\textbf{0}	& \cdots& \textbf{0}&\textbf{K}^*_R &  \textbf{K}_R
 \end{bmatrix}  \]
 which correspond to the dynamics of the left, interface, and right lattice respectively. These matrices are written as
}
\begin{widetext}
\[
		\textbf{K}_L = \begin{bmatrix}
			\hat{k}	& -k_2 			& \cdots 	& \cdots&  0 			\\
			-k_2 	& \hat{k} 		& -k_1		& \cdots &  0			\\
			0		&	-k_1 		& \hat{k}	& -k_2	& 0			\\
			\vdots	&\vdots			&\ddots 	&\ddots& \ddots					\\
			0 		&\cdots			&\cdots 	&-k_2&\hat{k}
		\end{bmatrix}, \ \
		\textbf{K}_I = \begin{bmatrix}
		\textbf{0}&	-k_1 & \hat{k} & -k_1&\cdots  & \textbf{0}\\
		\textbf{0}&\cdots &	-k_1 & \hat{k} & -k_2 & \textbf{0}
	\end{bmatrix}, \ \
		\textbf{K}_R = \begin{bmatrix}
			\hat{k}	& -k_1 			& \cdots 	& \cdots&  0 			\\
			-k_1 	& \hat{k} 		& -k_2		& \cdots &  0			\\
			0		&	-k_2 		& \hat{k}	& -k_1	& 0			\\
			\vdots	&\vdots			&\ddots 	&\ddots& \ddots					\\
			0 		&\cdots			&\cdots 	&-k_1&\hat{k}.
		\end{bmatrix} \ \ 
\]
\end{widetext}
where $\hat{k} = k_1+k_2+\gamma_0$ and the entries $\textbf{K}_L^* = [0,\cdots,0,-k_1]^\intercal$ and $\textbf{K}_R^* = [-k_2,0,\cdots,0]^\intercal$ account for the interaction between the left and right lattices with the interface.
The matrices $\textbf{K}_L$ and $\textbf{K}_R$ preserve the chiral symmetry of the system upon removal of the diagonal~\cite{Chaunsali2021,Suesstrunk2016}, e.g.,
\begin{equation}
	\begin{aligned}
		\Sigma_z[\textbf{K}_L -\textbf{I}_n(2+\gamma_0)]  +[\textbf{K}_L -\textbf{I}_n(2+\gamma_0)] \Sigma_z = \textbf{0}\\
		\Sigma_z[\textbf{K}_R -\textbf{I}_n(2+\gamma_0)]  +[\textbf{K}_R -\textbf{I}_n(2+\gamma_0)] \Sigma_z = \textbf{0}
		\label{Eq:chiral}
	\end{aligned}
\end{equation}
where $\Sigma_z$ is generated by the third Pauli matrix, $\sigma_z$. Note that the dynamical matrix of the full interface system, $\textbf{K}$, breaks the chiral symmetry only at the interface and will return a zero matrix plus a single non-zero entry of value $k_1-\hat{k}$ when evaluated with Eq.~\eqref{Eq:chiral}.

In the limit $\Gamma\to0$, which is $\bm{f}_{\rm NL}\to0$, (or equivalently in the limit of low energies), linear eigenvalue analysis can be performed on the linear system and the topological invariant of the bulk modes can be recovered analytically by imposing a plane wave solution $\bm{u}_j = \bm{u}(\bm{\mu})\exp(i\omega t + i\bm{\mu} j)$
where $\bm{u}(\bm{\mu})$ is the periodic Bloch eigenfunction of the infinite system parameterized by the wavevector $\bm{\mu}$.
The topological invariant which warrants the existence of the topological mode is \textcolor{edit1}{the winding number or equivalently} the Zak phase; this is a special case of Berry's phase for 1D lattices~\cite{Zak1989}. 
The Zak phase of the $m$-th band is expressed as the integral of the Berry connection over the Brillouin zone (BZ), 
\begin{equation}
	\theta_{\rm{Zak}} = \int_{-\pi}^{\pi}\left[i\bm{u}_{m}^\dagger(\mu)\cdot\partial_\mu\bm{u}_{m}(\mu)\right] \rm{d}\mu
	\label{eq:ZakPhase}
\end{equation}
where the superscript $(\cdot)^\dagger$ denotes the Hermitian transpose. 
In practice, this is numerically computed as~\cite{Xiao2015}
\begin{equation}
	\theta_{\rm{Zak}} = -\text{Im} \sum_{n=-N}^{N-1}\ln\left[\bm{u}_m^\dagger\left(\frac{n\pi}{N}\right)\cdot \bm{u}_m\left(\frac{n+1}{N}\pi\right)\right].
	\label{eq:ZakPhaseNumeric}
\end{equation}
\textcolor{edit1}{For $\gamma\neq 0$, either the left of right lattice will have a winding number $\nu = 1$ or equivalently $\theta_{\text{Zak}} = \pi$ indicating a nontrivial system while the adjacent lattice will remain trivial.
Looking just at the left lattice, the physical interpretation of this results is as follows. The left lattice undergoes a band transition when $\gamma$ turns from negative to positive and it's eigenfunctions acquire a quantized phase during an adiabatic evolution of the wavevector when $\gamma>0$, e.g.,~the optical branch switches phase while traversing the BZ. 
With a nontrivial invariant recovered for our system, the conditions for topological insulation are satisfied (see ref~\cite{Pal2018} for details on the infinite interface lattice). 
}

\begin{figure}[t!]
	\centering
	\includegraphics[width=\linewidth]{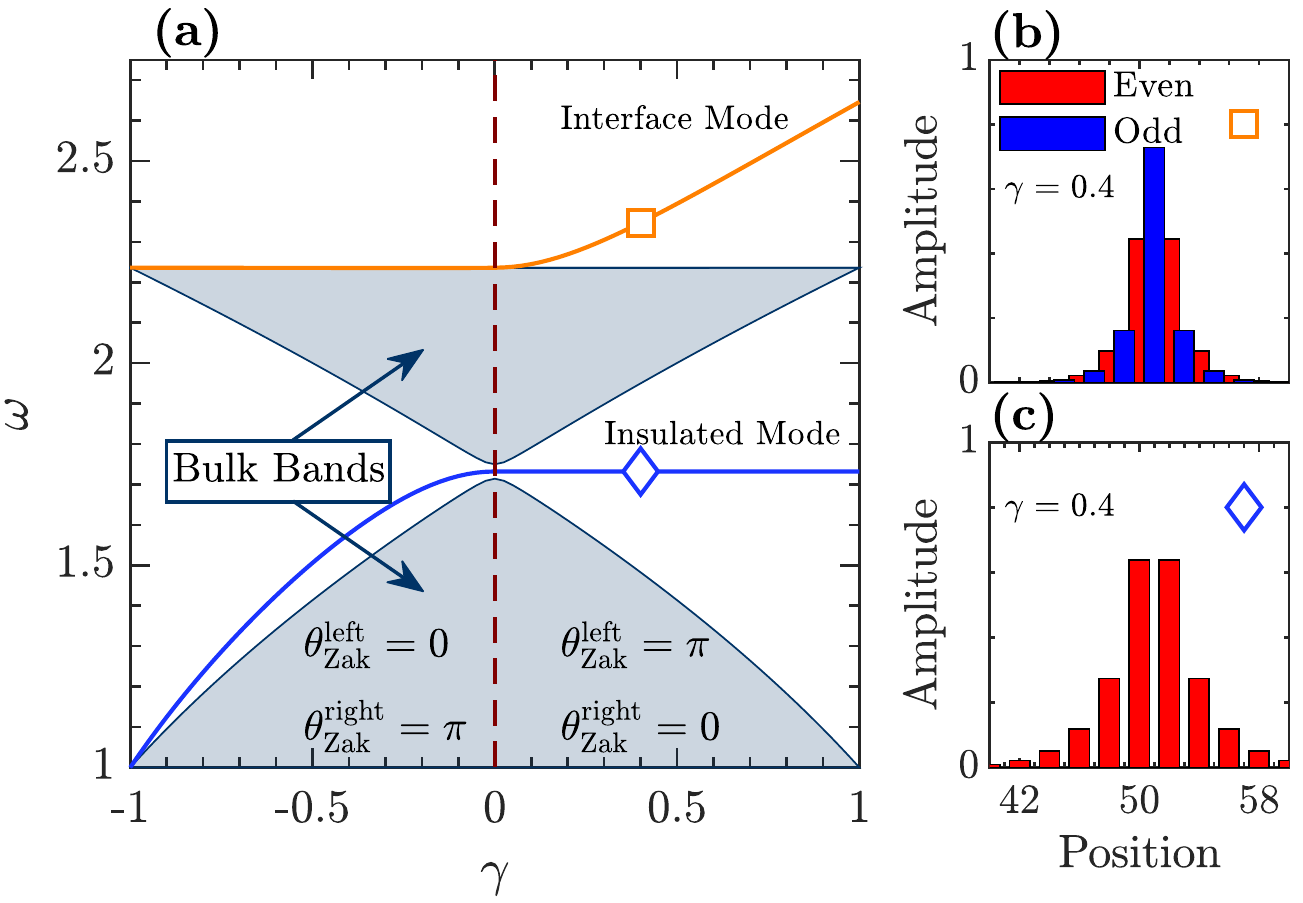}
	\caption{Linear eigenvalue analysis of the topological lattice (low-energy limit). The linear frequencies are shown (a) where a band-gap is present for $\gamma\neq 0$, with an inversion of the bands as  $\gamma$ increases from negative to positive, as indicated by the Zak phase. The linear normal modes for the interface mode (b) and topologically insulated mode (c) are shown as well for $\gamma = 0.4$. The odd lattice sites of the insulated mode remain stationary while the even sites oscillate at a linear frequency of $\omega = \sqrt{3}$. 
	}
	\label{Fig:Toplogical_Transition}
\end{figure}

Figure~\ref{Fig:Toplogical_Transition} displays this topological transition for the finite system as $\gamma$ evolves from -1 to 1. The corresponding eigenvalue problem of the linear finite system is
\begin{equation}
	\omega^2\textbf{X} = \textbf{KX}
\end{equation}
where $\textbf{X} = [u_1,u_2,\dots,u_n]^\intercal$. 
\textcolor{edit1}{When $\gamma<0$, chiral symmetry of the system is broken and the insulated mode produces a symmetric wave which is not centered in the band gap whereas 
	for $\gamma>0$, chirality is preserved resulting in an anti-symmetric insulated mode at the fixed linear frequency $\omega=\sqrt{3}$, and with it an interface mode emerges above the optical band~\cite{Chen2018,Pal2018,Vila2019}.}
The insulated mode is the result of the bulk-boundary correspondence. The linear interface mode follows an exponential amplitude decay away from the interface with out-of-phase motion between all oscillators, while the insulated mode displays the same amplitude decay but warrants displacements only at even lattice sites.

\subsection{Numerical Continuation} 
\label{sec:Numeric_Continuation}
\textcolor{edit1}{Now we address the effects of the strong nonlinear coupling $\Gamma$ on the dynamics of the topological lattice of Figure~\ref{Fig:Schematic}.}
Most previous work on nonlinear topological systems have relied on asymptotic methods to study the effect of the nonlinearity~\cite{Chaunsali2019,Hadad_2016,Pal2018,Narisetti2010}, however these are only applicable when the nonlinearity is weak so that the solutions can be regarded as small perturbations of the linearized ones.
Methods such as complexification averaging~\cite{Manevitch1999,Mojahed2019} have been developed for certain cases of strongly nonlinear or essentially nonlinear systems, however the application of this particular method is restricted to approximating the slow flow evolution component of dynamics.
In our system, we have a strongly nonlinear lattice for which asymptotic techniques generally do not apply. Thus, we follow a similar methodology to~\cite{Chaunsali2021, Vila2019} and make use of the method of numerical continuation to study the NNMs of the system. By this term, we denote nonlinear standing wave solutions of the system where all coordinates oscillate in a synchronous fashion, in similarity to linear vibration modes~\cite{Vakakis1996}.

The continuation technique applied in this work follows the methodology outlined in~\cite{Peeters2009}. 
The \textit{shooting method} initializes the system in state form $\dot{\bm{z}}  = f(\bm{z})$ where $\bm{z} = [\bm{u},\dot{\bm{u}}]^\intercal$ is the state vector. 
The first iteration initializes the NNMs with linear system analysis which is equivalent to the nonlinear normal modes at very small energy levels ($\mathcal{O}(10^{-5})$  for our system). 
The trial solutions of the dynamical system initiated at $\bm{z}_0$ are denoted by $\bm{z}(t,\bm{z}_0)$, and if the initial conditions satisfy a time-periodic evolution for some period $T$ then we denote them as $\bm{z}_{p}(t,\bm{z}_{p0})$ where $\bm{z}_{p}(t,\bm{z}_{p0}) - \bm{z}_p(t+T,\bm{z}_{p0}) = \textbf{0}$. 
The linear eigenvalue analysis will not perfectly satisfy condition of synchronicity required for a the NNM for finite energy levels, so set of initial conditions and period $(\bm{z}_{p0,(j)},T_{p0,(j)})$ are determined by using a Newton-Raphson scheme which iteratively minimizes the nonlinear function $\textbf{H}(\bm{z}_{p0,(j)},T_{p0,(j)}) = \bm{z}_{p}(\bm{z}_{p0,(j)},T_{p0,(j)}) - \bm{z}_{p0,(j)}$ until the initial conditions satisfy the tolerance condition,
\begin{equation}
	\frac{||\textbf{H}(\bm{z}_{p0,(j)},T_{p0,(j)})||}{ ||\bm{z}_{p0,(j)}||} < \varepsilon,
	\label{eq:Continuation}
\end{equation}
where $\varepsilon$ is a prescribed precision. This delivers the first NNM at low energy; the set of initial conditions and period $(\bm{z}_{p0,(j)},T_{p0,(j)})$ satisfying Eq.~\eqref{eq:Continuation} is the sought NNM at frequency $\omega = \frac{2\pi}{T}$ and associated energy $E$ for $j= 1$.
Using this known solution for the NNM, the next set of initial conditions is initialized using a \textit{predictor step} to formulate a guess of NNM $j+1$, $(\tilde{\bm{z}}_{p0,(j+1)},\tilde{T}_{p0,(j+1)} )$ which is generated along the tangent of the NNM branch in frequency-energy space at the point $\bm{z}_{p0(j)}$. The predictions are iteratively corrected until Eq.~\eqref{eq:Continuation} is satisfied for NNM $(j+1)$.
This continuation can be performed for any mode of the finite nonlinear lattice of Figure~\ref{Fig:Schematic} and results in an evolution in natural frequency and mode shape as a function of the system energy.

\subsection{Evolution of the Bulk-bands}
\label{sec:Bulk-Band Evolution}
\begin{figure*}[t!]
	\centering
	\includegraphics[width=\linewidth]{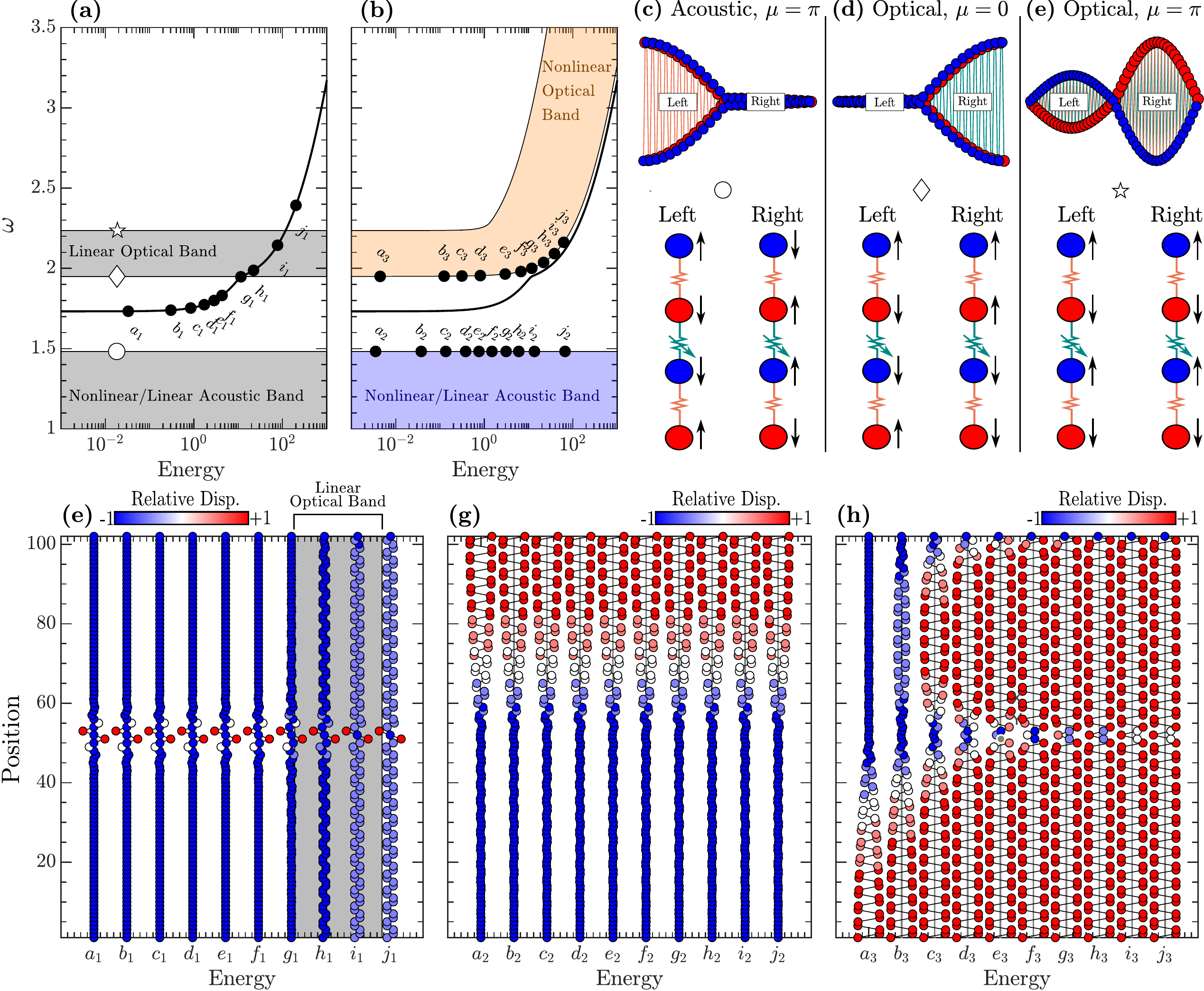}
	\caption{The evolution of the family of nonlinear normal modes associated with the topological insulation is shown to merge with the linear bulk-bands (a); in a more precise description of the dynamics, the nonlinear bulk-bands are shown to evolve with energy as well (b), meaning that the merger with the linear bulk-band as depicted in (a) is not meaningful at the high energy levels. 
	Also shown are the normal modes of the interface system in the low-energy limit, defining (c) upper boundary of the acoustic band, (d) lower boundary of the optical band, and (e) upper boundary of the optical band. Below the mode shape plots are schematics of even (red) and odd (blue) oscillators, with orange linear springs representing $k_2$ and green springs with a through-arrow representing $k_1+\Gamma$. These are provided for clarity to exemplify the in-phase versus out-of-phase motion of the modes at each band edge.	
		The mode shapes of the topological mode are altered as this merge happens (f), and the mode shapes of the bounding modes of the acoustic (g) and optical (h) bulk-bands are shown to evolve for only the optical branch.}
	\label{Fig:Mode_Evolution}
\end{figure*}

By performing the continuation technique described in section~\ref{sec:Numeric_Continuation}, we observe that for our system with grounding stiffness and inter-cell nonlinearity a frequency-energy dependency in both the topologically insulated mode and bulk-bands (cf.~Figure~\ref{Fig:Mode_Evolution}(a)).
In previous works on strongly nonlinear topological lattices, it was observed that the frequency-energy dependence of the nonlinearity causes a shift in the topological mode frequency of the NNMs, either softening or stiffening, until the mode merges with the linear band (either optical or acoustic)~\cite{Vila2019,Chaunsali2021}.
Furthermore, the evolution of the NNMs for the topological mode shows that it is at this point that the exponential decay in amplitude from the lattice interface loses its form \textcolor{edit1}{(around point $g_1$ of Figure~\ref{Fig:Mode_Evolution}(f))}, and displacements at extended lattice sites must be maintained at a comparable magnitude of displacement to the interface to maintain a time-periodic solution.  
However, prior work has described this phenomenon in the context of the linear system which, as we will show, does not represent the true dynamics the nonlinear lattice as the nonlinearity induces frequency-energy dependence on the bulk-bands.

\textcolor{edit1}{
The merging of the \textit{nonlinear} topological mode with the \textit{linear} bulk-bands was observed in~\cite{Vila2019} and described as 
 the transition from a spatially localized mode to an extended bulk-mode occurring at approximately the same frequency at which the topological NNM enters the the linear bulk-spectrum.
}
This empirically appears to be a compelling predictor for determining the existence of the topological mode based on the continuation of the topological mode, however it does not uncover the full picture of the system's nonlinear lattice. 
In strongly nonlinear lattices, the bulk-bands are expected to evolve with energy just as the topological mode does. 
This has been theoretically predicted through the method of numerical continuation, and experimentally verified~\cite{Mojahed2019}. The frequency-amplitude dependence caused by the nonlinearity causes the bulk spectrum to shift in frequency, and nonlinear hardening springs, such as ones with the potential $\Gamma(\Delta)^3$ in Figure~\ref{Fig:Schematic} (where $\Delta$ denotes displacement and $\Gamma>0$),  do increase the pass-band frequencies of the lattice. \textcolor{edit1}{To study this phenomenon in the topological system of Figure~\ref{Fig:Schematic}, we evolve the NNMs of the bulk-bands to explore (1) if the frequency-energy evolution of the nonlinear bulk-bands sheds any further light on the interpretation of the topological mode \textit{merging} with the bulk spectrum and (2) to present the full dynamics of the frequency-energy space of the topological system.}

The computation of the nonlinear bulk-band evolution with energy is performed as follows. 
The linear normal modes recovered from eigenvalue analysis in the low-energy limit are ordered in frequency. The bounding modes of the acoustic and optical bands are easily identified as the normal modes associated with the band-edge frequencies of the finite system. 
Furthermore, the band edges display certain qualitative behavior that aids in identifying their corresponding normal modes. 
For instance, the top-edge of the acoustic band displays in-phase behavior for all oscillators coupled by the stiff spring ($k_1$) and out-of-phase behavior for all oscillators coupled by the weak spring ($k_2$) (cf.~Figure~\ref{Fig:Mode_Evolution}(c)).  
Figure~\ref{Fig:Mode_Evolution} also depicts the linear mode shapes af the optical band. The in-phase and out-of-phase behavior of the optical band varies depending on whether or not the left (nontrivial) or right (trivial) half of the interface system is considered. For the left lattice, the bottom-edge of the optical band is characterized by in-phase oscillations for stiff couplings and out of phase couplings for soft couplings (just as for the acoustic band), while the right lattice is characterized by out-of-phase oscillations for stiff couplings and in-phase oscillations for weak couplings. Lastly, the top-edge of the optical band corresponds to out-of-phase oscillation between all oscillators. Note that this transition in phase behavior in the optical band is the physical manifestation of the nontrivial Zak phase and is a resemblance of the inversion the eigenfunctions of the infinite system.

The linear normal modes in the low-energy limit that define the boundaries of the linear bulk-bands (Figure~\ref{Fig:Mode_Evolution}(c-e)) are used to generate the nonlinear bulk-bands using the same numerical continuations scheme that was performed for the topological mode. The evolution of the band-edges envelopes the modified bulk-bands at each energy level. Figure~\ref{Fig:Mode_Evolution} shows the evolution of the bulk-bands as well as the evolution of the nonlinear normal modes that bound it. Note that since the stiff coupling elements (the coupling with nonlinear components $\Gamma$) do not affect the upper bounding in the linear normal modes of the acoustic band (cf.~Figure~\ref{Fig:Mode_Evolution}), the numerical continuation scheme computes normal modes free of nonlinear influences at high energy levels (a result which is verified in section~\ref{sec:Full System Analysis}). On the contrary, the bounding modes of the optical band are influenced by the stiffening nonlinearity since it raises their frequencies as the energy level increases and the linear limit of the system does not apply anymore.

Based on this methodology we arrive to the following result. 
\textcolor{edit2}{
The NNM which evolves from the nonlinear insulated mode approaches the lower edge of the nonlinear optical band as energy rises, and then follows the frequency-energy evolution of the optical band for high-energy oscillations. This is in contrast to the interpretation that the mode merges with the linear optical band, and this observation highlights the importance of calculating the band structure evolution with energy in nonlinear systems.%
}
\textcolor{edit1}{
However, there is a clear notching of the frequency-energy curve of the NNM based on the topological mode which coincides with the linear optical band frequency. 
\textcolor{edit2}{
Moreover, we find that at this point the topological NNM begins to decentralize from the interface, and that the mode disperses throughout the entire bulk while crossing through the linear optical band.%
}
 Thus, it is clear that the linear spectrum imputes influence over the evolution of the topological NNM. Nevertheless, it is around this frequency that the optical band evolves as well which prevents the aforementioned merging of the topological NNM with the bulk-spectrum at strongly nonlinear amplitudes.
}
Moreover, despite following very similar trajectories in the FEP, the NNM shapes of the bottom-edge of the optical band and the corresponding shape of the topological mode are opposites of one-another, that is, the oscillations in the bulk-band are dispersed across the extended bulk of the lattice while the oscillations of the topological mode remain concentrated at the interface. 
Note however that at sufficiently large energy levels, the topological mode will correspond to increasingly large amplitudes of lattice sites away from the interface until the standing wave (NNM) represents that of the optical band; however such energy levels require non-physical displacement amplitudes and are therefore not considered in this work.

\section{Stability Analysis of Standing Wave Solutions}
\label{sec: Stability Analysis}
We now consider  the stability of NNMs of the nonlinear lattice by using the method of Floquet Multipliers (FMs), which has already proven useful for studying the linear stability of one-dimensional topological lattices~\cite{Chaunsali2021}. The FMs of this analysis can be used to determine stability, and their argument can recover the Krein signature, a method which can indicate the nature of instabilities that occur in the nonlinear lattices~\cite{Marin1998,Panda2005,Flach2008}.

\begin{figure*}[t!]
	\centering
	\includegraphics[width=\linewidth]{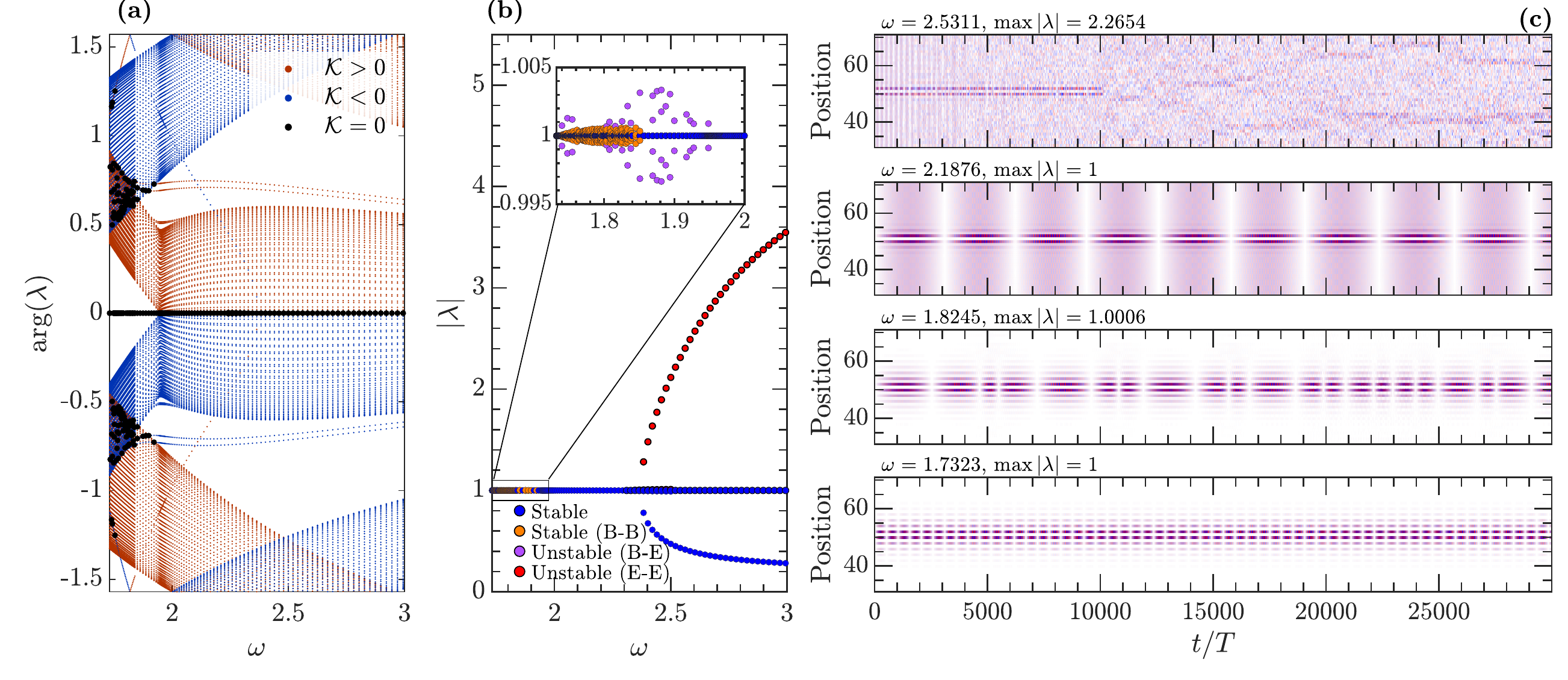}
	\caption{Krein signatures of the Floquet multipliers of the insulated mode (a) where blue represents $\mathcal{K}=1$ and orange $\mathcal{K}=-1$.  Also shown are the absolute values of the insulated mode's Floquet multipliers (b) with linearly stable, effectively stable via bulk-bulk (B-B) collisions, bulk-edge (B-E) collisions, and edge-edge linearly unstable (E-E) differentiated by color. The numerical simulations shown in (c) verify the stability, B-B stability, and E-E instability, and verify a high-energy linearly stable solution ($\omega = 2.1876$).}
	\label{Fig:Krein}
\end{figure*}

\subsection{Floquet Analysis}
\label{sec: Floquet Ananylsis}
Floquet theory can be used to recover linear stability of the standing wave solutions (NNMs). To this end, in state space, the system is expressed as 
\begin{equation}
	\dot{\bm{z}} = F(\bm{z})
	\label{eq:statespace}
\end{equation}
where $\bm{z}$ is again the state vector. The system is linearized by introducing the a small perturbation $\bm{z} = \bm{z}_0 + \bm{y}$, such that 
\begin{equation}
	\dot{\bm{y}} = \frac{ \partial F(\bm{z})}{\partial\bm{z}}\Bigg|_{\bm{z}_0} \bm{y} + \mathcal{O}\left(||\bm{y}||^2\right)
	= \textbf{A}(t)\bm{y}
	\label{eq:linearization}
\end{equation}
where $\bm{z}_0$ satisfies time-periodic solutions of Eq.~\eqref{eq:statespace}. This results in a linear time-periodic system, e.g.,~with period $T$, $\textbf{A}(t) = \textbf{A}(t+T)$. The stability problem is thus reduced to studying the zero solution of Eq.~\eqref{eq:linearization}. By standard Floquet theory, the solution of the perturbation at any time can be mapped from the initial state through the fundamental solution matrix $\textbf{C}(t)$ by $\bm{y}(t) = \textbf{C}(t)\bm{y}_0$. After $m$ periods of $t$, we have the relations 
\begin{equation}
	\bm{y}(mT) = \textbf{C}(mT)\bm{y}_0 = \textbf{C}(T)^m\bm{y}_0
	\label{eq:fundemental_mapping}
\end{equation}
where $\textbf{C}(T)$ is the monodromy matrix. For a solution to be bounded in time, we require that $\textbf{C}(T)^m$ be bounded, which is ensured for if each eigenvalue $\lambda_i$ of $\textbf{C}(T)$ satisfies the condition $|\lambda_i|\leq 1$. 
The eigenvalues of the monodromy matrix are the FMs of the system, and each corresponds to an eigenvector $\bm{v}_i$.

 Using numerical integration of Eq.~\eqref{eq:linearization}, the FMs are computed by evaluating the growth of the state vector after $m$ periods,
\begin{equation}
	\frac{|\bm{y}(mT)|}{|\bm{y}_0|} = |\lambda|^m
	\label{eq:numeric_floquet}
\end{equation}
Given that the nonlinear lattice of Figure~\ref{Fig:Schematic} is Hamiltonian, the symplectic structure of the monodromy matrix delivers FMs in sets of four which are complex conjugates and reciprocal complex conjugates of one-another~\cite{Marin1998}. 

\subsection{Krein Signature Analysis}
\label{sec:Stability of Standing Waves}

The eigenvectors that correspond to the FMs can be used to study the Krein signatures of the lattice, as has been done for a similar system with grounding nonlinearity~\cite{Chaunsali2021}. The Krein signatures are computed as,
\begin{equation}
	\mathcal{K}(\lambda_j) = {\rm{sign}}\left(-i\bm{v}^\dagger_j\begin{bmatrix}
		\textbf{0}_n & \textbf{I}_n \\ -\textbf{I}_n &\textbf{0}_n
	\end{bmatrix}\bm{v}_j\right),
\end{equation}
where $2n$ is the length of the state vector and $i=\sqrt{-1}$. The Krein signatures are quantized at values of $-1,0$, or $1$ corresponding to the sign of the eigenvector's energy. When spectral bands collide, e.g.,~an eigenvector branch of $\textbf{C}(t)$ with $\mathcal{K} = 1$ merges with one for which $\mathcal{K} = -1$, instabilities can occur in the system. 
The Floquet stability analysis is conducted on the nonlinear interface lattice at each iteration of the continuation scheme. 
Figure~\ref{Fig:Krein} shows the Krein signatures of the FMs as well as the absolute value of the FMs at each frequency step. 
\textcolor{edit1}{
Three types of instability are observed, namely bulk-bulk (B-B), bulk-edge (B-E), and edge-edge (E-E) instabilities. 
}

\textcolor{edit1}{ For $\omega<2$ we both B-B and B-E instabilities are shown in Figure~\ref{Fig:Krein}(a).}
\textcolor{edit2}{
The bulk-bulk instabilities occur when extended bulk-states coalesce~\cite{Chaunsali2021,Marin1998}.}
These instabilities appear in orange in Figure~\ref{Fig:Krein}(b) and are re-entrant, e.g.,~the FMs that depart the unit circle due to the collision rapidly fall back to the unit circle. The result is a very small instability that occurs over a narrow frequent range and is typically negligible in numerical simulations~\cite{Marin1998,Kopidakis1999}. Thus, we consider the B-B instability to be effectively stable for the sake of numerical experiments.
Note that the bulk-bulk instability is the result of the finiteness of the nonlinear lattice considered in Figure~\ref{Fig:Schematic}, since as the number of oscillators grows the total number of bulk-bulk instabilities is expected to increase but with smaller amplitudes. 
\textcolor{edit1}{
The other instability appearing for $\omega<2$ are B-E instabilities which appear in purple in Figure~\ref{Fig:Krein}(b). The Bulk-Edge instabilities are often referred to as \textit{internal modes}~\cite{Chaunsali2021,Flach2008} and the onset of B-E instabilities brings about a much larger rise in the FM leading to a numerically noticeable instability. 
}

\textcolor{edit1}{
The alternative E-E instability appears for $\omega>2.3$ and can be seen as the result of two spectral edge bands merging together on the real axis at $\arg(\lambda) = 0$ at $\omega\approx2.35$. 
These instabilities bring about much larger FMs which appear in red in Figure~\ref{Fig:Krein}(b).
The E-E instabilities occur when a branches of opposite-signed bulk-spectra bifurcate and intersects with each other. Because the E-E instabilities collide on the real axis, these are real instabilities as opposed to the oscillatory (Krein) instabilities observed in the B-B and B-E cases. Note that the B-B, B-E, and E-E instabilities correspond to black dots ($\mathcal{K}=0$) in Figure~\ref{Fig:Krein}(a), and we note that the FMs on the real axis are always zero by definition.
In between the B-E and E-E instabilities, we observe a nonlinear frequency regime where no instabilities are present. Thus, we arrive to the conclusion that was previously presented for a topological system with grounding nonlinearity~\cite{Chaunsali2021} for our interface system with coupling nonlinearity: The stiffening nonlinearity can support stable NNMs based on the linear topological edge state at high energy levels. 
}

To further study the stability of the computed NNMs, we numerically simulate the finite topological interface system of Figure~\ref{Fig:Schematic}. The initial displacements and velocities of the lattice system are chosen as state vectors satisfying the time-periodic solution at each given frequency. A perturbation of 1\% magnitude with respect to the initial conditions is applied, and the system oscillates for 30,000 periods. The numerical integration is carried out using the MATLAB$^{\text{\textregistered}}$ ODE45 routine. 

Figure~\ref{Fig:Krein}(c) depicts the results of the NNM stability analysis. The FEP shows the maximum FMs of the topological mode where the transitions from stability to instability are apparent, and the center plot shows the extremes of the FMs magnitudes at each frequency step. Four regimes are simulated, namely the low energy linear regime ($\omega = \sqrt{3}$), the bulk-bulk instability regime ($\omega = 1.8245$), the stable high-energy regime ($\omega = 2.1876$), and the strongly unstable regime ($\omega = 2.2654$). As expected, the low-energy linear regime produces a standing wave (NNM) localized at the interface, which supports oscillations at only the even lattice sites; this wave remains stable and does not lose its form for the duration of the simulation. The bulk-bulk instability is shown to be too insignificant to effectuate instabilities in the simulation, and a stable standing wave of the same profile as the linear topological wave exists indefinitely at the shifted frequency. The high-energy stable standing wave does not localize explicitly at the interface, as predicted by Figure~\ref{Fig:Mode_Evolution}, however it remains stable and is physically realizable provided that the corresponding initial conditions are selected. Lastly, the strongly unstable solution is shown as initial periodic motions localized at the lattice interface, but the energy quickly sheds into the surrounding lattice sites as instabilities dismantle the unstable standing wave solution.

\section{Geometric Phase Interpretation}
\label{sec:Geomteric Phase}
\textcolor{edit2}{
While the previous sections provide insight into the evolution and stability of standing waves (NNMs), the question of how the energy evolution of the system influences the topology of the bulk-bands requires further analysis. This is of great importance since the topological state is only excitable if the bands remain nontrivial.
}
\textcolor{edit1}{
\textcolor{edit2}{
It has been previously observed that the merging of the topological NNM with the linear bulk-spectrum can define the energy at which the topological mode is no longer excitable~\cite{Vila2019}. 
In this section, we seek to define a similar energy threshold using an alternative approach based on the transient dynamics of the bulk-bands and the observable results of the nontrivial band topology. 
}
As a starting point, we seek numerical realizations of the geometric phase in the frequency-energy space of the lattice. 
}

\subsection{Numerical Realization of Zak Phase}
\label{sec:Numerical Zak Phase}
%

The geometric phase of interest is the Zak phase, which is a special case of Berry's phase for 1D periodic systems~\cite{Zak1989}. Its formulation is well-established for linear systems (see section~\ref{sec:NumericalContinuation and Nonlianer Pass Bands}), however, its interpretation in strongly nonlinear systems is not yet accomplished~\cite{Zhou2020} since this issue is typically addressed only for the case of weak nonlinearity~\cite{Chaunsali2019,Pal2018,Hadad2018}. 
While the Berry phase (and Zak Phase) can be generalized for nonlinear systems, the result is not typically quantized and is therefore not a suitable topological invariant~\cite{Tuloup2020}.
The recent work by Zhou \textit{et. al.}~has addressed this by offering an analysis procedure for the strongly generalized nonlinear Schrodinger equation~\cite{Zhou2020}. The methodology used to quantize the nonlinear Berry phase draws its reasoning by invoking results of the symmetry types of the nonlinear lattice considered and by analyzing the trajectories (orbits) of the derived solutions. Here, we use this methodology to quantify the Zak phase of our nonlinear lattices.

The prescribed methodology outlined in~\cite{Zhou2020} is as follows.
A nonlinear bulk-mode for a dimer lattice at wavenumber $\mu=q$  assumes a solution of the form 
\begin{equation}
	\bm{u}_q = \left[u^{\text{even}}_q(\omega t- qn), u^{\text{odd}}_q(\omega t- q n + \phi_q) \right]^\intercal
	\label{eq:NL_bulkmode}
\end{equation}
where $\phi_q$ is the phase difference between even and odd lattice sites at wavenumber $\mu = q$. 
If reflection symmetry is obeyed, $(u_n^{\text{even}},u_n^{\text{odd}}) \to (u_{-n}^{\text{odd}},u_{-n}^{\text{even}})$, then the bulk-mode corresponding to $\mu = -q$ is also a solution to Eq.~\eqref{eq:NL_bulkmode}, e.g.~$ \bm{u}'_{-q} = \left[u^{\text{odd}}_q(\omega t- qn), u^{\text{even}}_q(\omega t- q n + \phi_{-q}) \right]^\intercal$. 
If the system is not degenerate, this implies $ \bm{u}'_{-q} =  \bm{u}_{-q}$, so $\phi_{-q} = -\phi_q\mod2\pi$. Thus, the geometric phase can be captured by evaluating $\phi_q$ at the high-symmetry points of the Brillouin zone (BZ), namely when,
\begin{equation}
	\theta_{\text{Zak}} = \phi_\pi -\phi_0.
	\label{eq:phase_NL}
\end{equation}

For an adiabatic evolution of a wavenumber $\mu$ traversing the BZ, a geometric phase is acquired if the phase between oscillators changes between high symmetry points ($\mu=0$ and $\mu = \pi$). Evaluating the phase condition between even and odd lattice sites is sufficient to return the geometric phase for our nonlinear lattice system in Figure~\ref{Fig:Schematic}. 
\textcolor{edit1}{
This treatment is of a similar procedure to experimentally recovering Zak phases in~\cite{Xiao2015}, and provides the required physical interpretation of the nontrivial phase condition.
}

\subsection{Numerical Experiments on Half-Lattices}
\label{sec:Numerical Experiements half lattice}

\begin{figure*}[t!]
	\centering
	\includegraphics[width=\linewidth]{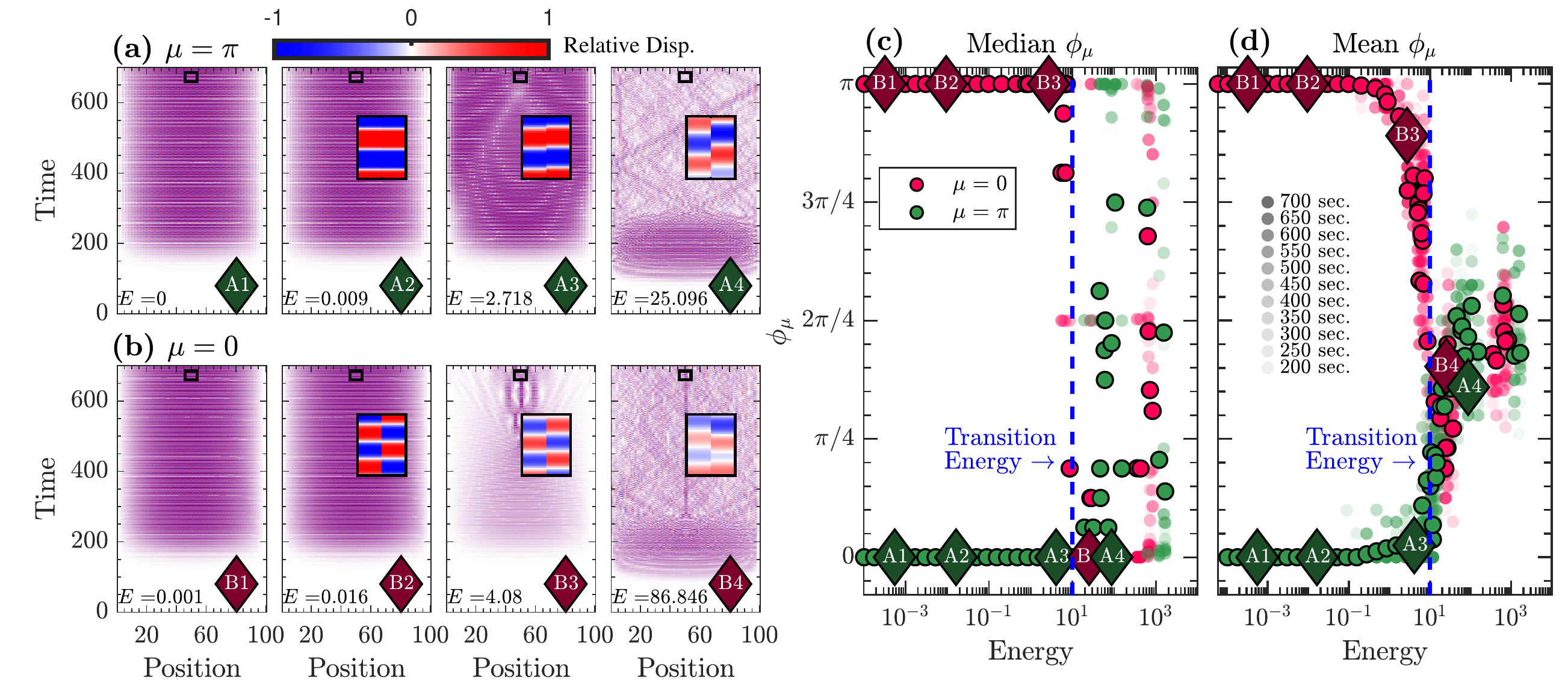}
	\caption{The phase difference between even and odd lattice sites for the left (nontrivial) half-lattice. The numerical simulations of the optical branch are shown in for (a) $\mu=0$ and (b) $\mu = \pi$ to display the mode-symmetry inversion at the high-symmetry points of the Brillouin zone. The median (c) and mean (d) phase $\phi_\mu$ both become ill-defined at approximately $E=12$, as denoted by the vertical dashed line. In order to empirically observe the Zak phase, the phase difference between even and odd lattice sites must be different; it follows that this dashed line is a predictor regarding the critical energy level where the full nonlinear system of Figure~\ref{Fig:Schematic} will no longer satisfy the conditions for the existence of a topologically protected mode. The phase conditions of several snapshots in time are also plotted in (c,d) to show the convergence behavior of these values where time is indicated by the transparency value of the small markers}
	\label{Fig:PhaseComp}
\end{figure*}

\textcolor{edit1}{To empirically investigate the band topology of our nonlinear system, we isolate the left (nontrivial) lattice to study the geometric phases of its bulk-modes with no interface. 
%
Because our system is strongly nonlinear, the traditional Bloch theorem can not be applied and the finite system must be studied
A nonlinear lattice system is considered with $n=102$ degrees of freedom (51 unit cells) of the left lattice (nontrivial) configuration. In practice, the modes will be excited via external harmonic forcing, so we apply harmonic excitation to excite the lattice rather than prescribe initial conditions. 
A harmonic excitation is applied in the form of a Gaussian tone burst until a prescribed energy level is achieved, resulting in the following equations of motion,}
\begin{equation}
	\ddot{\bm{u}}+\textbf{Q}\bm{u}+\bm{f}_{\rm NL}  = 
	\begin{cases} 
		\bm{\eta} S \exp\left({i\omega_{\rm ext} t - \frac{(t-t_0)^2}{\tau^2}}\right), & E< E^* \\
		\textbf{0}& E\geq E^* \\
	\end{cases}
	\label{eq:forcing}
\end{equation}
\textcolor{edit1}{%
\textcolor{edit2}{%
where $\bm{\eta}$ is a vector which profiles the relative magnitude and phase of forcing at each location, $\omega_{\rm ext}$ is the excitation frequency, $\tau = 25T$ where $T$ is excitation period, $t_0 = 5\tau$, $\bf{Q}=\bf{K}_L$, and $S$ is the amplitude which varies depending on $E^*$ which is the desired energy level of the lattice for a given simulation.}
 Here, the vector $\bm{\eta}=\textbf{X}_i$ denotes the modal displacement of eigenvector $i$ of the finite half-lattice system. The values of $i$ associate with the modal displacement of the optical band at either $\mu=0$ or $\mu = \pi$. This profile of forcing ensures that the bulk-mode is excited.
 Two excitation frequencies are considered to correspond to the eigenvectors for $\mu = 0$ and $\mu = \pi$. Namely, these are the frequencies that define the lower and upper boundaries of the optical pass band. 
}
 These frequencies correspond to the high-symmetry points of the lattice, and excitation at these prescribed frequencies will reveal the phase conditions which the lattice assumes at steady state. 

Figure~\ref{Fig:PhaseComp}(a) displays the simulation results for a harmonic excitation that matches the frequency of the lower high symmetry point ($\mu = 0$) of the optical band of the left lattice at increasing levels of energy. Likewise, Figure~\ref{Fig:PhaseComp}(b) displays the same simulation results, but for excitation frequencies of the upper high-symmetry point ($\mu = \pi$). By the definition of $\phi_q$, we have the condition $\phi_q= 0$ if there is in-phase motion between even and odd lattice sites for $\mu = q$, and $\phi_q = \pi$ if the corresponding motion is out-of-phase.  It is clear form Figures~\ref{Fig:PhaseComp}(a) and (b) that at $\mu = 0$, the phase is $\phi_0 = 0$ and that $\phi_\pi = \pi$   for sufficiently low energies; hence, by the definition of Eq.~\eqref{eq:phase_NL}, we recover the anticipated results (by linear theory) that $\theta_{\rm{Zak}} = \pi$ indicating a nontrivial topology. When energy levels rise, however, the wave motion is distorted by the nonlinearity, and these phase conditions become increasingly ambivalent.

To quantify the phase conditions at various energy levels, the Fast-Fourier-Transform (FFT) is employed to decompose the time histories of the simulations. The phases of the individual lattice sites are tracked using the argument of the resulting spectra, and the differences between even and odd sites return the geometric phase conditions. Figure~\ref{Fig:PhaseComp}(c-d) shows the numerical results. 
\textcolor{edit1}{
The mean and median phase conditions of all dimer sites (51 in total) are plotted together, and an ensemble of samples are used for each frequency at various temporal locations in the simulation as denoted by the marker transparency of Figures~\ref{Fig:PhaseComp}(c) and (d).}\textcolor{edit1}
{The stability of the phase conditions at each recorded time provides assurance that the measured phase conditions are indicative of the systems stead-state dynamics once excitation has ceased.}
Interestingly enough, the phase differential between the high-symmetry points makes a clear transition at a critical energy level which is approximately $E = 12$. Considering the median values, the upper band no-longer holds the consistent median value of $\phi_\pi = \pi$ at this energy level, and this is the same point at which the mean values of $\phi_0$ and $\phi_\pi$ coincide with one another. 
Accordingly, a vertical line marks the critical energy level at which the harmonic forcing produces waves with no distinct geometric phase, and the lattice loses its topologically nontrivial condition. It is precisely at this point when one may predict that the full interface system will no longer support an externally excited topologically insulated wave.
\textcolor{edit1}{
	In juxtaposition to the NNM analysis of the previous section, here we have employed the observable products of nontrivial band topology to define an energy threshold for which the mode existence based on theoretical results of topological band theory.
}
 This is theoretical predication based in the numerically computed Zak phase will be confirmed by the numerical simulation results discussed in the next section. 

\section{Full System Analysis} 
\label{sec:Full System Analysis}

The full nonlinear lattice system is studied numerically for the same excitation profile given by Eq.~\eqref{eq:forcing}, with 102 degrees of freedom; namely, 25 unit cells (50 oscillators) at the left lattice, 1 interface unit cell (2 oscillators), and 25  unit cells (50 oscillators) at the right lattice. 
\textcolor{edit1}{Here, forcing is only applied at the interface to excite the topologically protect mode, e.g., $\bm{\eta}= \delta_{(1+n/2,i)}$, and $\bf{Q}=\textbf{K}$.}
The force is applied to lattice site 52, which is intended to excite the topological mode that supports even-site displacements explicitly. The same excitation parameters as the previous section are employed for the full system analysis ($\tau = 25T$, $t_0 = 5\tau$) unless otherwise specified.

\subsection{Numerical Experiments of Full System}
\label{sec:full system numeric}

\begin{figure*}[t!]
	\centering
	\includegraphics[width=\linewidth]{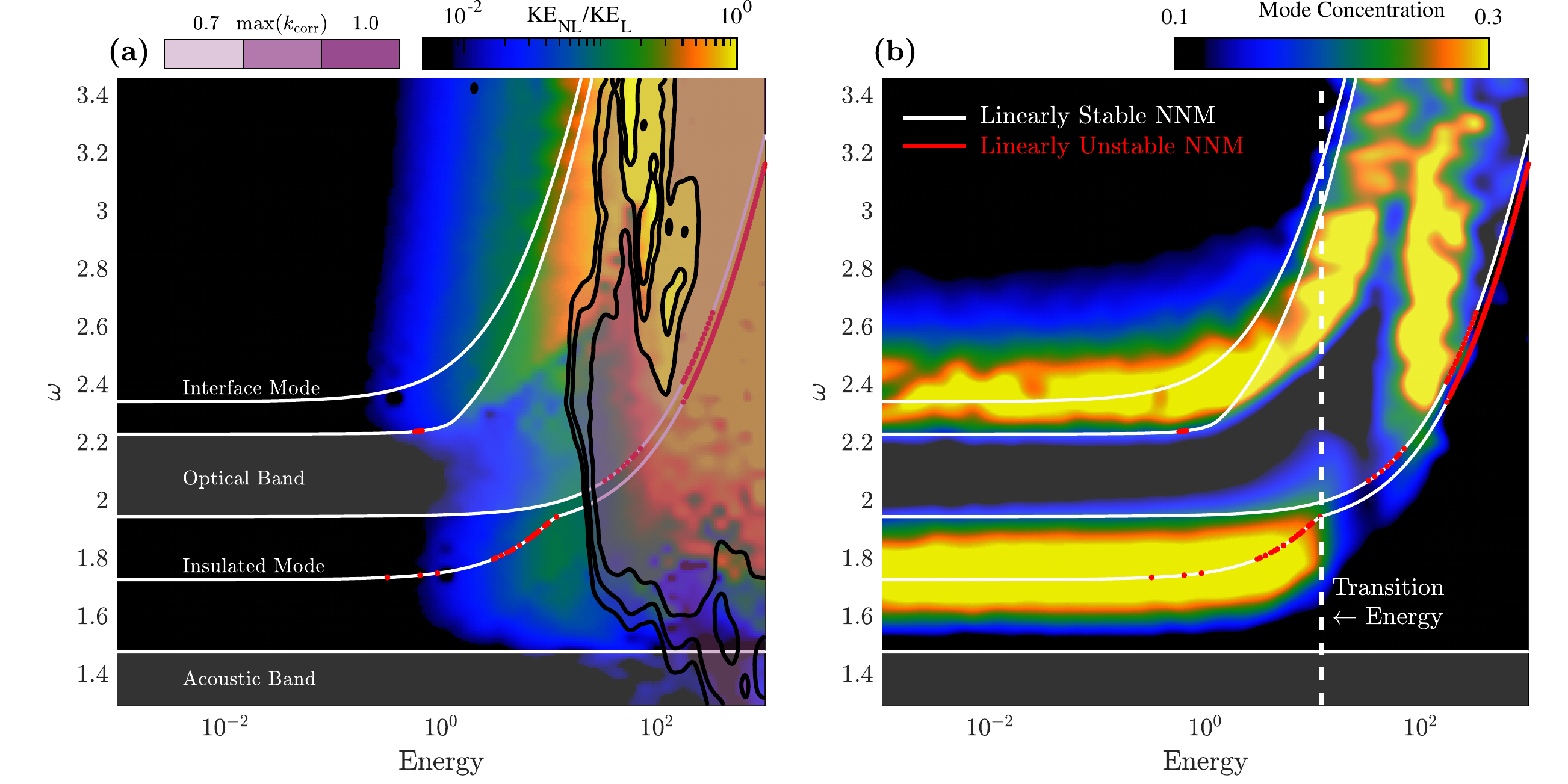}
	\caption{The ratio of total kinetic energy of nonlinear to linear potentials as well as the boundaries for chaotic dynamics is shown (a) next to the mode concentration analysis conducted over a range excitation energies and frequencies (b). 
		The excitation is applied with a Gaussian signal until the prescribed energy level is reached. A $k_{\rm{corr}}$ score greater than 0.7 is a strong indicator that chaotic dynamics are present in the simulation. The FEP is overlayed onto the mode concentration plot to show the evolution of the nonlinear lattice's optical pass band with the predicted bounding nonlinear normal modes. The red segments on the FEP correspond to NNMs which were found to be linearly unstable. The vertical dashed line marks the $E = 12$ energy level, which as predicted based on empirical phase observations the topological mode will no longer exist. This is confirmed by the sudden drop in mode concentration for excitation regimes which follow the nonlinear continuation of the insulated mode at this energy level. }
	\label{Fig:FEP_ModeConentration}
\end{figure*}

A grid of 2500 iterations of excitation frequencies and energy levels is considered as the parametric domain of the following numerical study. The frequency values vary linearly in the range $\omega\in(\sqrt{3}/2, 3\sqrt{3}/2)$, and the energy values vary in logarithmic steps in the range $E\in(10^{-3},10^{3})$. These ranges ensure that the frequency-energy space of the FEP which encompass the nonlinear normal modes of the bulk spectrum, the topologically insulated mode, and the topological interface mode are probed (as predicted by Figure~\ref{Fig:Mode_Evolution}). The instantaneous energy of the topological lattice at any time instant is computed as
\begin{equation}
	E = \frac{1}{2}\dot{\bm{u}}^\intercal\dot{\bm{u}} + \frac{1}{2}{\bm{u}}^\intercal\textbf{K}{\bm{u}}  + E_{\rm NL}
\end{equation}
 whereas the nonlinear energy component $E_{\rm NL} = \sum_{2i\leq N/2}\frac{1}{4}\Gamma(u_i - u_{i+1})^4 + \sum_{ N/2 \leq(2i+1)\leq N}\frac{1}{4}\Gamma(u_i - u_{i+1})^4 $  accounts for the nonlinear couplings between positions $u_i$ and $u_{i+1}$.  Once the prescribed energy ratio is reached, the forcing is set to zero.

Our aim is to uncover which frequency-energy pairings result in a topologically insulated mode at the interface of the lattice at steady state  (e.g.,~after many periods after the excitation ceases). 
The harmonic force typically requires between 50 and 400 seconds to generate the prescribed energy level (with higher energy simulations requiring more forcing), so a simulation time of 1000 seconds is used to ensure that the steady-state dynamics has been reached. 
To quantify the localization of the oscillations (e.g.,~whether or not the energy remains localized to a subset of oscillators), we employ the same modal concentration ratio ($MCR$) given in~\cite{Chaunsali2021},
\begin{equation}
	MCR = \frac{\sum_{i=1}^{N} u_i^4}{\left(\sum_{i=1}^{N} u_i^2\right)^2},
	\label{eq:modal_concentration}
\end{equation}
where $u_i$ is the displacement of mass $i$. The $MCR$ of each simulation is computed for the root-mean-square value of the lattice displacements from the time when the force ceases to the end of the simulation. This time-averaging of oscillator displacements captures the amplitudes of both stationary and traveling waves in the lattice.

Figure~\ref{Fig:FEP_ModeConentration} shows the results of this numerical experiment with the FEP of the nonlinear normal modes overlayed. A vertical white line at the critical energy level of $E = 12$ marks the energy level at which the numerical geometric phase experiments predicted that the nonlinear insulated mode would no longer be supported (see results of section~\ref{sec:Geomteric Phase}). To the left of the $MCR$ plot is the FEP with the linear stability of the five important NNMs marked by the line color.

\begin{figure}[t!]
	\centering
	\includegraphics[width=\linewidth]{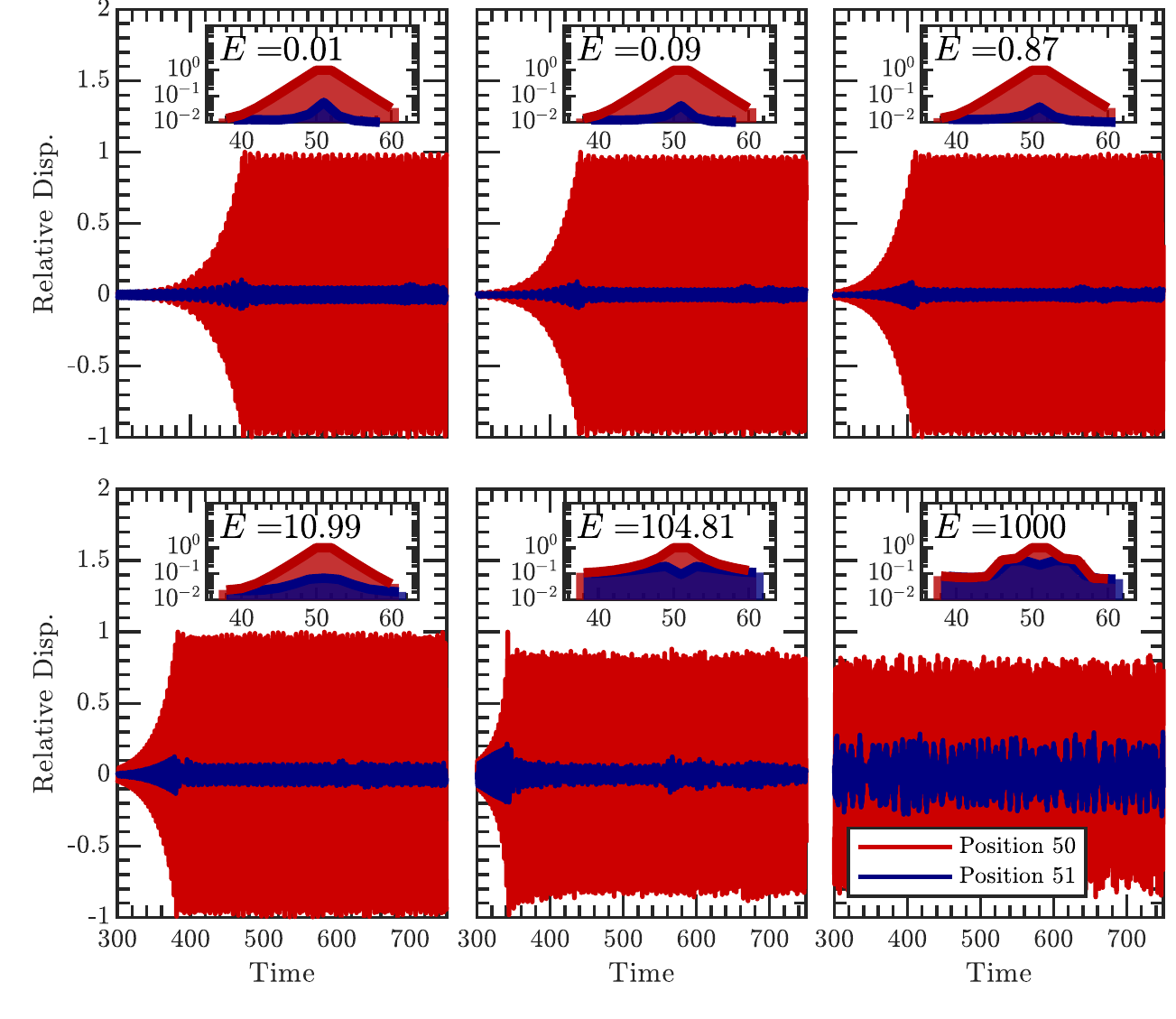}
	\caption{Time histories of the simulations for no perturbations at the interface location. The excitation frequencies match those of the family of topologically insulated modes at the prescribed energy level. The top right corner of each plot displays the steady state root-mean-square amplitude of each lattice site on a logarithmic scale with even sites appearing in orange and odd sites appearing in purple, and the energy level of each simulation.}
	\label{Fig:ModeTS}
\end{figure}

Starting with the bulk-bands, it is apparent that the simulations with forcing within the acoustic-band frequencies support the FEP predictions for the wave to propagate freely though the lattice with little dependence on energy. The more interesting bulk-band is the optical pass-band which is predicted to have frequency-energy dependence at its bounding  NNMs. In the (low-energy) linear limit, the optical pass bands of the numerical experiment are clearly bounded within the theoretical prediction as indicated by the lower $MCR$ values. Once sufficiently high energy levels are reached, the optical pass-band clearly increases in frequency with amplitude and follows the nonlinear pass-band predictions of the continuation analysis. 
\textcolor{edit1}{
To ensure that strong nonlinear effects are achieved in this study, the average potential energies of nonlinear components and linear components are computed for each simulation. Figure~\ref{Fig:FEP_ModeConentration} shows that around $E = 1$, the nonlinear effects become prominent; eventually the nonlinear dynamics overtake the linear dynamics as energy is increased.}

\textcolor{edit1}{
At sufficiently large energies ($E\geq 100$), the nonlinearity in the system produces chaotic dynamics and causes localized oscillations irrespective of topological properties within the optical band as indicated by the overlayed contours shown in Figure~\ref{Fig:FEP_ModeConentration}(a). Here, the 0--1 binary diagonstic for chaos which was presented in~\cite{Gottwald2009} was used to define such a threshold by using the time-histories of the simulations. The 0--1 method was selected over the more-traditional Lyapunov exponent analysis because the numerical tools presented by Wolf~\cite{Wolf1985} or Rosenstein~\cite{Rosenstein1993} for resolving the Lyapunov spectrum require a proper state-space reconstruction; selecting the appropriate parameters for this procedure is not trivial for a finite lattice with open boundary conditions.
In contract, the 0--1 method does not make any assumption about dimensionality or state-space structure and only requires that the dynamics are deterministic. 
The theoretical underpinnings of the 0--1 test can be found in~\cite{Gottwald2016}. In brief, the method presented in~\cite{Gottwald2009} uses the projection of the time-series $\phi(t)$ to define data in a new $p$-$q$ plane via $p(n) = \sum_{j=1}^n \phi(j)\cos (jc)$, $q(n) = \sum_{j=1}^n \phi(j)\sin (jc)$ where $c\in(0,\pi)$, and then uses correlation measures between the covariance of the time-series and  the mean-square displacement between $p(n)$ and $q(n)$ to arrive at a score $k_{\text{corr}}$. In practice, many realizations of $p(n)$ and $q(n)$ are considered, and the median value is taken as $k_{\rm{corr}}$ in order to avoid possible resonance between $\phi(t)$ and $c\in(0,\pi)$.  A score near the value 1 indicates strongly chaotic motion, while a score near 0 indicates regular periodic motion. To arrive at the contours shown in Figure~\ref{Fig:FEP_ModeConentration}(a), the $k_{\text{corr}}$ score was computed for the time-histories of each oscillator, and the maximum score was returned. Note that we implemented the necessary sub-sampling scheme to avoid the false-positives that may be returned~\cite{Gottwald2016,Melosik2016}.
}


\textcolor{edit1}{
Looking now at the topologically insulated mode, it is apparent that the insulated mode is excited as indicated by the high $MCR$ values that follow the insulated NNM at low energies.} To confirm that the mode shape of the insulated mode is recovered, Figure~\ref{Fig:ModeTS} plots the first 250 seconds of the time-histories for various energy levels excited at a frequency selected from the nonlinear normal mode generated from the linear insulated mode. In the upper right corner, the root-mean-square amplitudes are plotted on a logarithmic scale confirming that the corresponding displacements follow an exponential decay from the interface, with motion occurring mainly in even oscillators. 

Figure~\ref{Fig:FEP_ModeConentration}(b) shows that the nonlinear topologically insulated mode appears to continue to follow the frequency-energy dependence of the FEP up until a sudden drop in the $MCR$ value. This drop occurs very near to the $E = 12$ energy mark which the phase analysis predicted to be the critical energy level where the topological mode ceases to exist. 
\textcolor{edit1}{
This is approximately the same energy level at which the NNM of the insulated mode converges to the NNM shape which defines the lower edge of the nonlinear optical pass-band, and this interestingly occurs at approximately the same frequency at which the topological NNM enters the frequency regime of the linear optical band; this further supports the importance of the optical band in predicting the topological mode evolution.%
}
\textcolor{edit2}{%
The B-B and B-E instabilities recovered from Floquet analysis do not demonstrate any discernible relationship between linear stability of the NNM and the excitability of the topological mode, however this is not a surprising result since FMs only apply to NNMs.
%
\textcolor{edit2}{%
Thus, we can deduce the following. The numerical continuation of the system's NNMs provide a reliable predictor for the frequency-energy evolution of the topological mode and the bulk-bands. The notching, which ensues when the frequency of the topological NNM is within that of the linear bulk-spectrum, defines the same energy thresholds which are found from empirical observations of the bulk-band topology, and this threshold is confirmed by numerical simulations of the topological system.
}
 }

\subsection{Influence of Forcing Profile on the Topological Mode}
\label{sec:Influence of forcing profile}

\begin{figure}[t!]
	\centering
	\includegraphics[width=\linewidth]{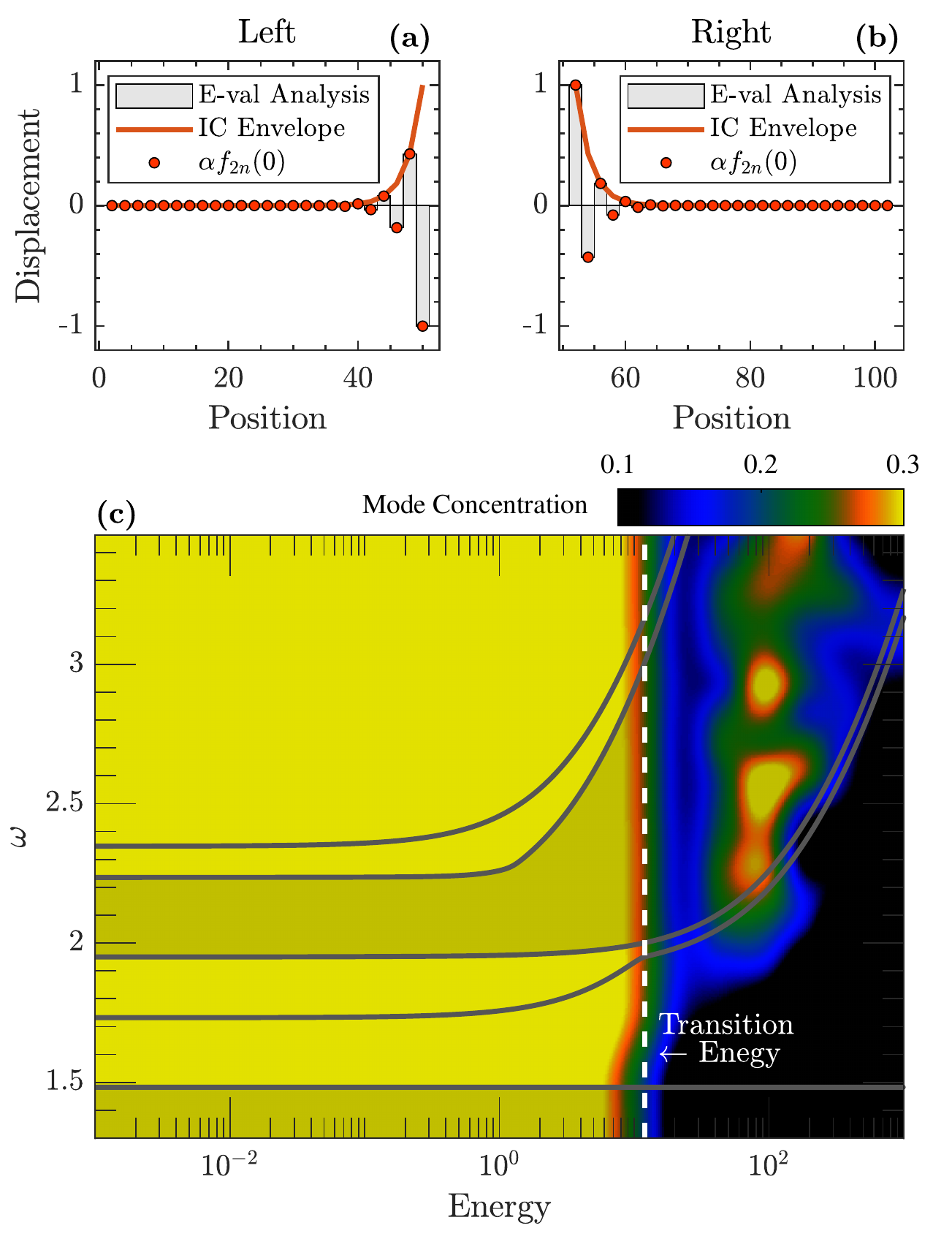}
	\caption{Topologically insulated mode results for forcing profiles which mimic the linear topological mode shape scaled to amplitude $\alpha$. The exponential decay of amplitude (a,b) prescribe an exponential forcing envelope, with the excitation at each site is governed by Eq.~\eqref{eq:forcing} until the prescribed energy level is reached. Using this forcing regime, the topological mode is excited at all frequency levels up until the $E = 12$ energy point as shown by the mode concentration ratios (c). This is due to the fact that the edge mode is perfectly realized, and as a result the lattice is restricted to oscillate in its linear limit until the nonlinearity influences the systems dynamics at higher energy levels. }
	\label{Fig:exponentialforcing}
\end{figure}

Because the nonlinear topologically insulated mode follows an exponential decay in amplitude from the interface site, it is of interest to determine what effect harmonic forcing away from the interface may play on its existence. To this end, a forcing envelope is employed that applies excitation to each lattice site in proportion to the site's amplitude as predicted by the (low-energy) corresponding linear insulated mode. 
\textcolor{edit1}{
The equations of motion are of the same form of Eq.~\eqref{eq:forcing} where now $\eta_n = ae^{bn}(-1)^{n/2}$ and again $\bf{Q}=\bf{K}$.
The parameters $a$ and $b$ fit the exponential envelope where $n = 2,4,\dots,N$ in order to only excite even lattice sites as prescribed by the linear eigenmode.} 

Figure~\ref{Fig:exponentialforcing} shows the $MCR$ values over the same grid that was constructed for Figure~\ref{Fig:FEP_ModeConentration}. When exciting the system with the exponential envelope, it is clear that the FEP analysis may be discarded as there is seemingly no frequency-energy dependence with regard to $MCR$ values. Rather, the existence of the topologically insulated mode depends solely on the level of energy provided to the system. It is found again that approximately at the critical energy threshold of $E=12$ (marked with a black vertical line) is the threshold for which the insulated mode exists. For energy levels beneath this critical level, it does not matter what frequency the system is excited at as the low amplitude oscillations of the bulk of the lattice supports oscillations that remain within the linear limit~\cite{Zhou2020}. When energy increases, however, the nonlinearity alters the dynamics of the system and this is no longer the case.

\subsection{Robustness to Perturbations}
\label{sec:Perturbations}
\begin{figure}[t!]
	\centering
	\includegraphics[width=\linewidth]{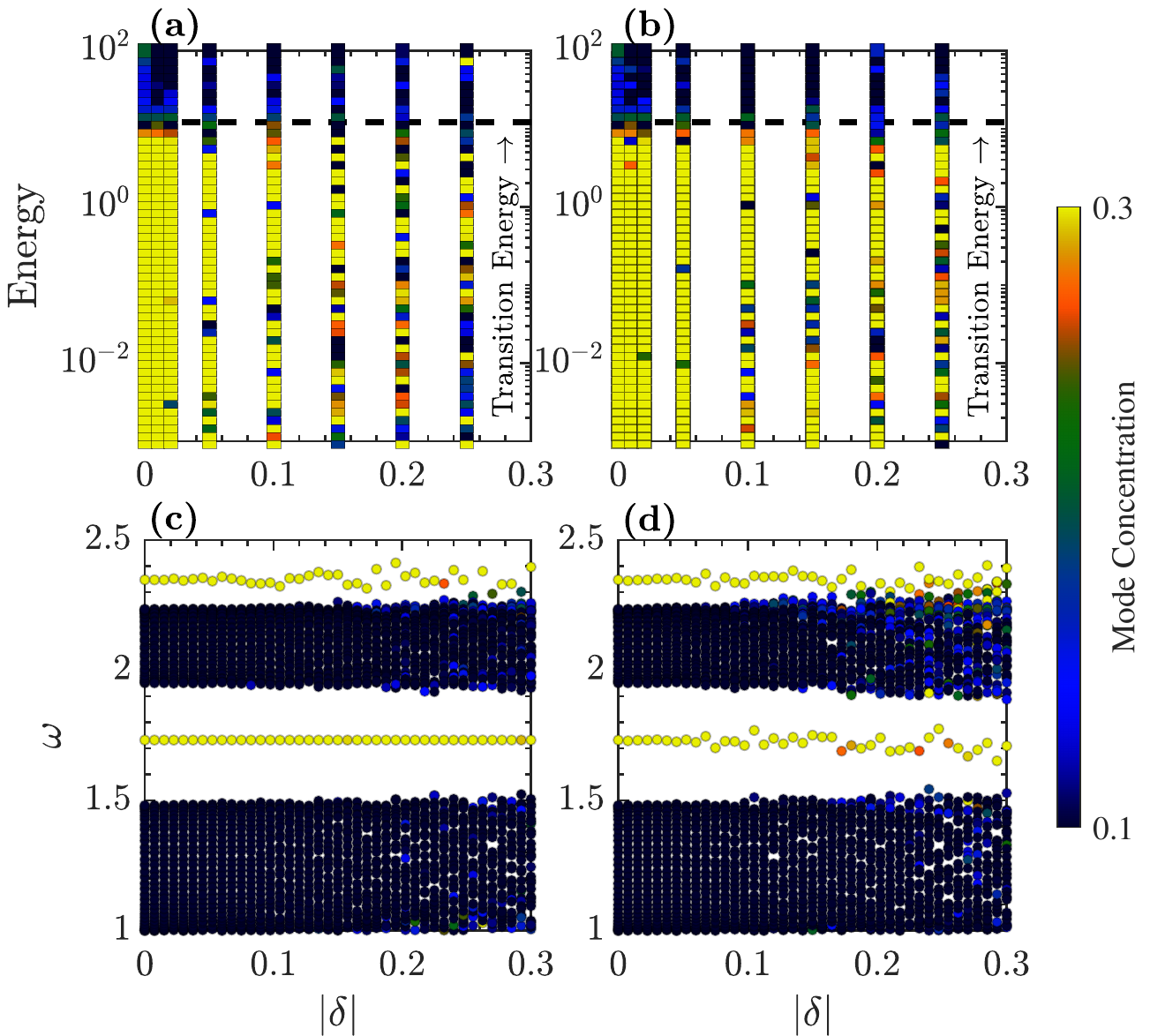}
	\caption{The effects of chiral (left) versus achiral (right) perturbations on the existence of the topological mode. Plots (a) and (b) show the $MCR$ values of the topological mode at each perturbation level for chiral and achiral perturbations respectively, while plots (c) and (d) show the eigenvalue analysis of the corresponding (low-energy) linear system for chiral and achiral perturbations respectively. \textcolor{edit1}{The value of $|\delta|$ shown on the axes indicates the average perturbation level of across the system.}}
	\label{Fig:perturbations}
\end{figure}

\begin{figure}[t!]
	\centering
	\includegraphics[width=\linewidth]{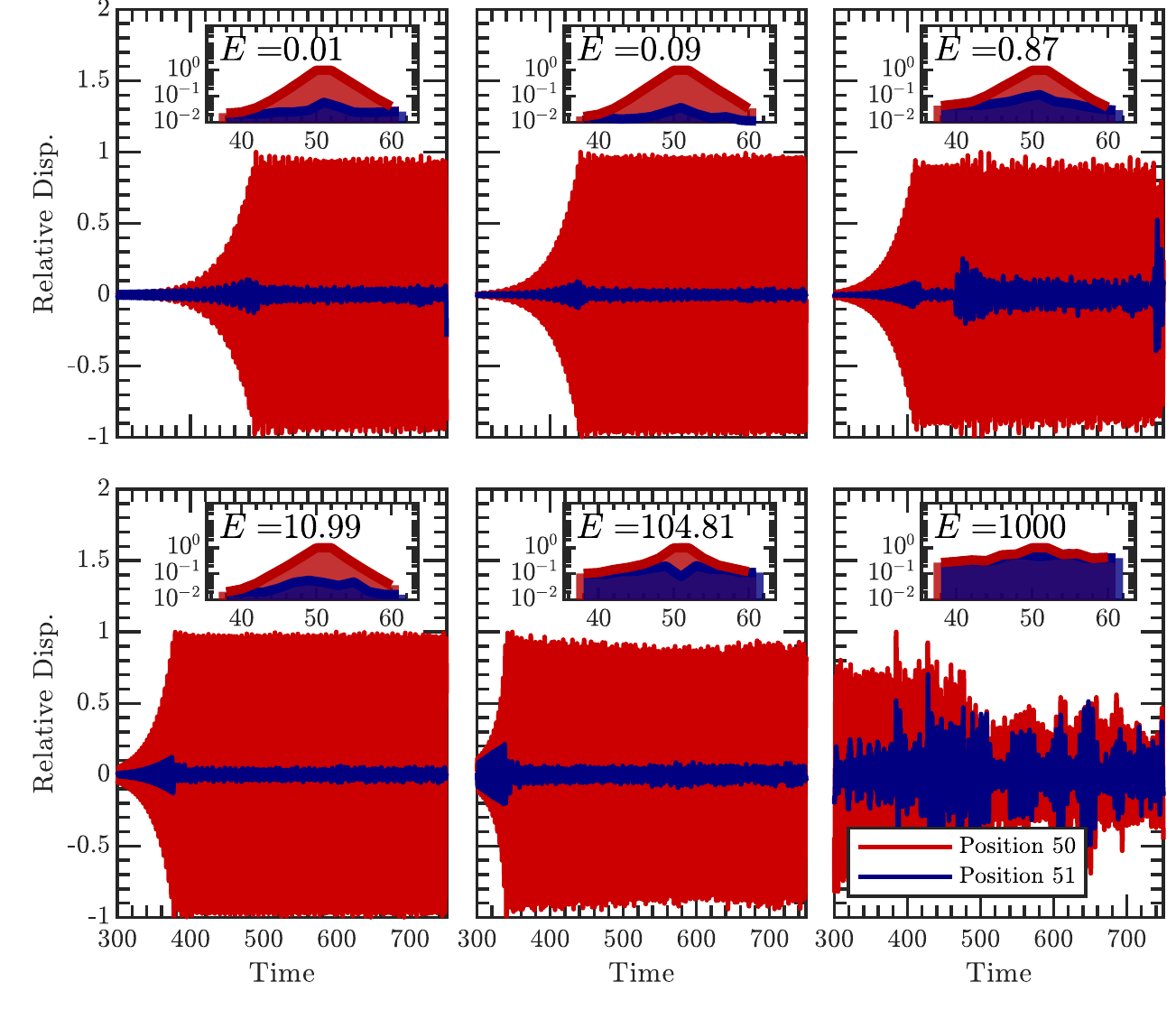}
	\caption{Time histories of the simulations for a perturbation level of $\delta = 0.05$. The excitation frequencies match those of the family of topologically insulated modes at the prescribed energy level. The top right corner of each plot displays the steady state root-mean-square amplitude of each lattice site on a logarithmic scale with even sites appearing in orange and odd sites appearing in purple, and the energy level of each simulation. }
	\label{Fig:perturbed_TS}
\end{figure}

One of the distinct features of topological insulators is their robustness to perturbations~\cite{Chaunsali2017}. To study this attribute in our system and numerical experimental scheme, we evaluate the $MCR$ of the insulated mode after subjecting the topological lattice of Figure~\ref{Fig:Schematic} to parametric perturbations of various magnitudes. Because the two half-lattices which make up the interface system preserves chiral symmetry, a symmetry type required for this type of insulator~\cite{Suesstrunk2016}, it may be anticipated that perturbations which preserve the chiral symmetric conditions of Eq.~\eqref{Eq:chiral} may alter the insulated mode differently compared to perturbations which are achiral as was shown in~\cite{Chaunsali2021} for the NNM solutions a their system. 
In contrast, we show that when exciting the insulated mode of the nonlinear interface lattice of Figure~\ref{Fig:Schematic} that while system in fact is robust to perturbations, the chirality of the perturbations are irrelevant to the result.

Chiral symmetric perturbations are applied such that the diagonal of the matrix $\textbf{K}$ does not deviate. The perturbation at each unit cell obeys $k_1 = k(1+\gamma) \to k(1+\gamma + \delta_1)$ and $k_2 = (1-\gamma) \to (1-\gamma+\delta_2)$, with the grounding term serving to return the diagonal to its nominal value, $\gamma_0 \to \gamma_0-\delta_1-\delta_2$. Alternatively, achiral perturbations assign a random variation of magnitude $\delta$ to each spring in the system.

Figure~\ref{Fig:perturbations} demonstrates the robustness of the topological system to both forms of perturbations.
Figures\ref{Fig:perturbations}(c) and \ref{Fig:perturbations}(d) depict the linear eigenvalue analysis of the nonlinear lattice dynamics at various levels of $\delta$ for chiral and achiral perturbations respectively. 
It is seen that chiral symmetric perturbations maintain the linear insulated mode frequency at $\omega = \sqrt{3}$, whereas the linear insulated mode shifts in frequency at roughly $\omega\propto\delta\sqrt{3}$ when the perturbations are achiral.
The system is forced at frequency levels predicted by the continuation of the topologically insulated mode. Figures~\ref{Fig:perturbations}(a) and \ref{Fig:perturbations}(b) show the recovered $MCR$ values for $\delta\in(0,0.25)$. For $\delta\to0$, the $MCR$ values match that of Figure~\ref{Fig:FEP_ModeConentration} for the insulated mode, as expected. 
As $\delta$ rises, there is a remarkable robustness of this result for both chiral and achiral perturbation schemes; the $MCR$ values do not appear to deviate from the unperturbed result until $\delta = 0.05$. 
Furthermore, the system preserves its ability to support the topologically insulated modes up to $\delta = 0.25$, but with far more variation of $MCR$ scores for low-energy states. As before, this mode is only present for $E<12$ J, as indicated by the vertical dashed lines. The consistency of the response in the perturbed system is confirmed by considering the corresponding time series at $\delta =0.05$ in Figure~\ref{Fig:perturbed_TS} which show an \textcolor{edit2}{acceptable match} to the time histories of Figure~\ref{Fig:ModeTS}. 
\textcolor{edit1}{
While this brief study should be performed over an ensemble of iterations before casting definitive conclusions about the exact behaviors of the system after undergoing perturbations, this would be computationally cumbersome to achieve and is outside of the scope of this work. Nevertheless, we include these results as a demonstration that the previous claims in this study are relatively impervious to parametric perturbations. 
}

\section{Conclusions}
\label{sec: conclusion}
We have analyzed a mechanical analog of the SSH interface lattice model with a cubic nonlinearity between stiff couplings. Two half lattices are interfaced, one with a nontrivial topological invariant and the other with a trivial topological invariant. In the linear (low-energy) limit, this lattice is shown to support a topologically insulated mode in the center of the system's bandgap. 
The effects of the nonlinearity on standing wave (NNM) solutions  are investigated by performing numerical continuation of the linear normal modes to find time-periodic solutions, or NNMs, at high energy states for both the topologically insulated mode and bulk-bands.
The stability of the NNMs is investigated by computing the Floquet Multipliers (FMs) and Krein signatures of the time-periodic solutions. 

Like the system explored in~\cite{Chaunsali2021}, it is shown that the nonlinear topological mode can maintain stable standing waveforms at high energy levels. 
While the FMs of predict stability between approximately 1.7 to 2.3 rad/s of frequency for the NNMs which evolve from the topologically insulated mode, it is shown that the NNMs evolve into standing waves which do not obey the exponential amplitude decay of topological modes for $\omega>1.95$, which is approximately the frequency of the lower edge of the linear optical band. 
Bulk-bulk, bulk-edge, and edge-edge type instabilities are found with Krein signature analysis. Numerical simulations show that the bulk-bulk instabilities do not disrupt the stability of the topological mode for 30,000 periods of oscillations (Figure~\ref{Fig:Krein}(c)), however the NNM does lose its stability in the regime of bulk-edge instabilities. 

\textcolor{edit2}{
By performing continuation on the bounding modes of the bulk-bands, we show the evolution of the nonlinear lattice's optical and acoustic bands with energy. Even though the topological mode loses its structure upon entering the \textit{linear} optical band, we show that in the framework of nonlinear bulk-bands that the nonlinear topological mode does not enter the \textit{nonlinear} optical band.}
Rather, it asymptotically approaches the evolution of the optical band over increasing energy and follows the boundary of the nonlinear optical band in the Frequency-Energy plot (FEP, Figure~\ref{Fig:Mode_Evolution}). 
\textcolor{edit1}{
However, we note that the linear bulk-band frequencies influence the evolution of the topological mode as seen by the notch in the topological NNM branch at approximately the linear pass-band frequency (Figure~\ref{Fig:Mode_Evolution}).}

To predict the existence of the topological mode under the influence of external excitation, numerical realizations of the Zak phase are performed by tracking the relative phase of even and odd lattice sites. This leads to the prediction that at approximately $E =12$, the lattices lose their nontrivial phase condition required for the topologically protected mode to exist.
\textcolor{edit2}{Interestingly, this transition occurs at the same energy level at which the notching appears in the topological NNM. Thus, we conclude that studying the geometric phases of the bulk-bands can be viewed as a transient-based and band theory motivated alternative to previous methods based on locating the intersection of the topological NNM and the linear bulk-spectrum.}
 Numerical simulations of the full system corroborate this claim as it is shown that the modal concentration ratio in the FEP space following the topological mode sharply declines beyond this point (Figure~\ref{Fig:FEP_ModeConentration}), confirming that the edge mode is recovered before this limit (Figure~\ref{Fig:ModeTS}). This analysis also supports the nonlinear bulk-bands predicted by the continuation analysis of the band-edges. Furthermore, it is shown that when the entire lattice is forced with an envelope proportional to the linear insulated mode, the same $E=12$ energy mark is found (Figure~\ref{Fig:exponentialforcing}). Lastly, the effects of perturbations are explored, and the topological mode is shown to be robust to chiral and achiral perturbations up to 5\% perturbation levels (Figure~\ref{Fig:perturbations}).

\textcolor{edit2}{
FM and Krein analysis are employed to uncover the regions of linear stability and instability in the FEP of the topological mode and bulk-bands. Numerical simulations confirm standing wave stability in the topological lattice with regard to various oscillatory instability types. However, no connection is discernible between the stability analysis of the NNMs and the excitability of the topological mode. Due to the fact that FM and Krein analysis are only applicable to NNMs, it is not surprising that no such relation is identified. Nevertheless, it would be interesting to further explore the relation between linear stability analysis of standing wave solutions and the existence of the topological mode excited by external forces in future studies.
}

Given the modular nature of mechanical systems, it may be possible in future work to confirm these results with experimental configurations. Additionally, future work  investigating other forms of nonlinearity (e.g.,~quadratic, Hertzian, etc.) could be explored and compared to the current findings, as well as the effect of making nonlinearity essential (e.g.,~possessing no linear stiffness component). Lastly, while this work has contributed to the understanding of 1D nonlinear topological systems, an extension of these methods to topological insulators of higher dimensions capable of supporting corner, surface, or edge states would be of great interest as they have primarily been studied for linear or weakly nonlinear couplings.

\section*{Acknowledgment}
This work was supported by the National Science Foundation Graduate Research Fellowship Program whose support is gratefully acknowledged.  
\bibliography{Topological_Protection_bib}

\end{document}